\documentstyle[amssymb,aps,floats,graphicx]{revtex}
\begin{document}
\title{RADIATION\ FROM\ ACCELERATED PERFECT OR DISPERSIVE MIRRORS FOLLOWING
PRESCRIBED RELATIVISTIC ASYMPTOTICALLY INERTIAL TRAJECTORIES}
\author{A Calogeracos *}
\address{Division of Theoretical Mechanics, Hellenic Air Force Academy TG1010,\\
Dhekelia Air Force Base, Greece\\
(*) acal@hol.gr}
\date{1 August 2001}
\maketitle

{\bf Abstract}

We address the question of radiation emission from both perfect and
dispersive mirrors following prescribed relativistic trajectories. The
trajectories considered are asymptotically inertial: the mirror starts from
rest and eventually reverts to motion at uniform velocity. This enables us
to provide a description in terms of {\it in }and {\it out }states. We
calculate exactly the Bogolubov $\alpha $ and $\beta $ coefficients for a
specific form of the trajectory, and stress the analytic properties of the
amplitudes and the constraints imposed by unitarity. A formalism for the
description of emission of radiation from a dispersive mirror is presented.

\section{\protect\bigskip Introduction}

The question of emission of radiation from imperfect mirrors has received
substantial attention in the last ten years. The present work is greatly
influenced by Barton and Eberlein (1993) where a Hamiltonian treatment of
radiation from a dielectric mirror is presented, by Jaeckel and Reynaud
(1993) and Barton and Calogeracos (1995) who treat dispersive mirrors, and
by Lambrecht, Jaeckel, and Reynaud (1996), Law (1995), Sch{\"{u}}tzhold,
Plunien, and Soff (1998) where radiation from vibrating cavities is
considered. The formalism has been exploited to study decoherence; see Maia
Neto and Dalvit (2000). There are two common threads in the above works: (i)
the treatments follow the Hamiltonian formulation of quantum field theories,
(ii) the calculations are nonrelativistic (notice however that the model
presented in Barton and Calogeracos (1995) is manifestly covariant). This
situation has to be contrasted with the study in the 70s of radiation from
relativistic perfect mirrors following prescribed trajectories (Davies and \
Fulling (1977) being one of the important early papers on the subject). In
this connection we should note that (a) perfect mirrors have not yet been
amenable to a Hamiltonian treatment, (b) the treatment of perfect mirrors
has traditionally focussed on the relation to black hole radiation. As a
result of (b) the trajectories considered were characterized by acceleration
nonvanishing in the $t\rightarrow \infty $ limit. We examine the latter
problem in a companion paper (A.\ Calogeracos (2001)), hereafter referred to
as II. Our objective here is to treat radiation from dispersive mirrors,
moving along prescribed relativistic asymptotically inertial trajectories.
The approach we adopt in this note may be of use in the calculation of the
radiation emitted by the interface between two media when its speed
approaches a characteristic speed of the system (cf the motion of the A-B
interface in He3).

The present paper is organized as follows. In section 2 the normal modes for
a one-sided perfect mirror are presented in the manner of Fulling and Davies
(1976), Davies and Fulling (1977), and Birrell and Davies (1982). A physical
interpretation of the modes in terms of the lab frame is given. In section 3
we consider a one-sided mirror that starts from rest, accelerates along the
trajectory $z=-\ln \left( \cosh t\right) $ till it reaches a space-time
point {\it P }where it has some arbitrary velocity $\beta _{P}${\it ,} and
then reverts to motion at uniform velocity. The particular trajectory is of
interest because it gives rise to a thermal spectrum if the acceleration is
to go on forever. We present an exact calculation of $\beta \left( \omega
,\omega ^{\prime }\right) $ in terms of hypergeometric functions, and
discuss the analytic properties of the $\alpha \left( \omega ,\omega
^{\prime }\right) $ and $\beta \left( \omega ,\omega ^{\prime }\right) $
amplitudes and the constraints imposed by unitarity. There are several
advantages in considering asymptotically inertial trajectories: (i)
Acceleration continuing for an infinite time implies mathematical
singularities and also entails physical pathologies associated, for example,
with the infinite energy that has to be imparted to the mirror. Notice that
the point {\it P} may lie as close to the light cone as desired, or in other
words the mirror's final speed may be close to the speed of light. (ii) The
mirror's rest frame eventually (after acceleration stops) becomes an
inertial frame and the standard description in terms of {\it in }and {\it %
out }states is possible. One may choose either the lab frame or the mirror's
rest frame to describe the photons produced. (iii) One avoids statements
about photons produced {\it while }the mirror is accelerated; rather one
makes unambiguous statements pertaining to times $t=\pm \infty $ when
acceleration vanishes. We show in general that for an asymptotically
inertial trajectory with the velocity being everywhere continuous the
amplitude squared $\left| \beta \left( \omega ,\omega ^{\prime }\right)
\right| ^{2}$ goes as $\left( \omega ^{\prime }\right) ^{-5}$ for large $%
\omega ^{\prime }.$ This is in contrast to a trajectory accelerated forever
where $\left| \beta \left( \omega ,\omega ^{\prime }\right) \right| ^{2}$
goes as $1/\omega ^{\prime }$ for large $\omega ^{\prime }$. This radical
difference is due to a subtle cancellation between two contributions, one
arising from the initial and the other from the final asymptotic part of the
trajectory; see II for details. In section 4 we treat dispersive mirrors
adopting the manifestly covariant approach of Barton and Calogeracos (1995).
The dispersivity of the mirror is controlled by a parameter $\gamma $. We
give the {\it in }and {\it \ out }modes for a dispersive mirror and outline
how the method of section 3 may be applied to calculate the spectrum of the
radiation emitted. In Appendix A we give some mathematical details on the
trajectory.

\section{The normal modes}

\subsection{The standard treatment}

Let us introduce coordinates 
\begin{equation}
u=t-z,v=t+z  \label{e01}
\end{equation}

\noindent The massless Klein-Gordon equation reads 
\begin{equation}
%TCIMACRO{\dfrac{\partial ^{2}\phi }{\partial u\partial v} }
%BeginExpansion
{\displaystyle {\partial ^{2}\phi  \over \partial u\partial v}}%
%EndExpansion
=0  \label{e1}
\end{equation}

\noindent Hence any function that depends only on $u$ or $v$ (or the sum of
two such functions) is a solution of (\ref{e1}). Let $z=g(t)$ be the
mirror's trajectory. We take everything to exist to the right of the mirror.
The objective is to find a set of modes satisfying (\ref{e1}) and the
boundary condition 
\begin{equation}
\phi (t,g(t))=0  \label{e2}
\end{equation}

\noindent One set of modes (Birrell and Davies (1982), eqn (4.43)) is 
\begin{equation}
\varphi _{\omega }(u,v)=%
%TCIMACRO{\dfrac{i}{2\sqrt{\pi \omega }} }
%BeginExpansion
{\displaystyle {i \over 2\sqrt{\pi \omega }}}%
%EndExpansion
\left( \exp (-i\omega v)-\exp \left( -i\omega p(u)\right) \right)  \label{e3}
\end{equation}

We will always think of (\ref{e3}) and also of (\ref{e5}) below as functions
of the lab coordinates $\left( t,z\right) $ through (\ref{e01}). The
determination of the function $p(u)$ goes as follows. For a certain value of 
$u$ consider the point where the line $u=${\it constant }cuts the mirror's
trajectory (figure 1). Let $t(u)$ be the time coordinate of that point.

Define 
\begin{equation}
p(u)=2t(u)-u  \label{e004}
\end{equation}

\noindent Since by (\ref{e01}) 
\begin{equation}
t(u)-u=g\left( t(u)\right)  \label{e04}
\end{equation}

\noindent it follows that 
\begin{equation}
p(u)=t(u)+g\left( t(u)\right)  \label{e4}
\end{equation}

\noindent From the second of (\ref{e01}) it follows that 
\begin{equation}
v=p(u)  \label{e05}
\end{equation}
It is clear from (\ref{e05}) that $\varphi _{\omega }(u,v)$ vanishes on the
trajectory. Equation (\ref{e05}) describes the trajectory in terms of the $%
u,v$ variables. Conversely 
\begin{equation}
u=f(v)  \label{e005}
\end{equation}
where the function $f$ is the inverse of $p$. The graphical interpretation
is that an outgoing (rightward) light ray corresponding to a certain $u$
cuts the trajectory at a point where the incoming (leftward) light ray is
labelled by $v=p(u)$. From definitions (\ref{e01}) and (\ref{e05}) it
follows that for points $(t,z)$ on the trajectory 
\begin{equation}
t+z=p(t-z)  \label{e005b}
\end{equation}

\noindent The following observation will be of use later on. Consider motion
along the trajectory $-g(t)$, i.e. the reflection of the trajectory $g(t)$
with respect to the origin. It is clear from (\ref{e005b}) that the
equations for the two trajectories are related via the correspondence $%
u\rightleftharpoons v,f\leftrightharpoons p$.

We introduce the velocity 
\[
\beta =\frac{dg}{dt} 
\]

\noindent Differentiating (\ref{e005b}) with respect to time we obtain 
\begin{equation}
p^{\prime }(u)=\frac{1+\beta (u)}{1-\beta (u)}  \label{205}
\end{equation}

\noindent where $\beta $ stands for the instantaneous velocity (not to be
confused with the amplitude $\beta (\omega ,\omega ^{\prime });$ the latter
will always be a function of two variables). From the fact that $f$ and $p$
are inverse it follows that 
\begin{equation}
f^{\prime }(v)=\frac{1-\beta (v)}{1+\beta (v)}  \label{206}
\end{equation}

For motion at uniform velocity $B$ the above expression yields 
\begin{equation}
f_{0}(v)=C+\frac{1-B}{1+B}v  \label{12b}
\end{equation}

\noindent where the constant $C$ is related to the initial condition. >From (%
\ref{205}) and (\ref{206}) we may readily obtain $\beta $ along the
trajectory as a function of either $u$ or $v$: 
\begin{equation}
\beta (u)=\frac{p^{\prime }(u)-1}{p^{\prime }(u)+1}  \label{207}
\end{equation}

\begin{equation}
\beta (v)=\frac{1-f^{\prime }(v)}{1+f^{\prime }(v)}  \label{208}
\end{equation}

\noindent In the case of the accelerating trajectory $g(t)=-\ln \left( \cosh
t\right) $ considered in section 3 \noindent the function $f(v)$ is denoted
by $f_{acc}$ and given by (\ref{a4}) (with $\kappa =1$): 
\begin{equation}
f_{acc}(v)=-\ln \left( 2-e^{v}\right)  \label{208b}
\end{equation}

The quantity $d\beta /du$ will turn up in the treatment of dispersive
mirrors. We write 
\begin{equation}
\frac{d\beta }{du}=\frac{d\beta }{dt}\frac{dt}{du}  \label{209}
\end{equation}

\noindent and from the definition of $u$ we get 
\begin{equation}
\frac{du}{dt}=1-\beta  \label{210}
\end{equation}

\noindent Combination of (\ref{209}) and (\ref{210}) yields 
\begin{equation}
\frac{d\beta }{du}=\frac{a}{1-\beta }  \label{211}
\end{equation}

\noindent where $a$ stands for the acceleration $d\beta /dt$. Similarly we
obtain 
\begin{equation}
\frac{d\beta }{dv}=\frac{a}{1+\beta }  \label{212}
\end{equation}

Notice that another set of modes satisfying the boundary condition is
immediately obtained from (\ref{e3}) 
\begin{equation}
\bar{\varphi}_{\omega }(u,v)=%
%TCIMACRO{\dfrac{i}{2\sqrt{\pi \omega }} }
%BeginExpansion
{\displaystyle {i \over 2\sqrt{\pi \omega }}}%
%EndExpansion
\left( \exp \left( -i\omega f(v)\right) -\exp \left( -i\omega u\right)
\right)  \label{e5}
\end{equation}

\noindent We quote the time derivative for future reference 
\begin{equation}
\frac{\partial \bar{\varphi}_{\omega }(u,v)}{\partial t}=\frac{\omega }{2%
\sqrt{\pi \omega }}\left( f^{\prime }\left( v\right) \exp \left( -i\omega
f(v)\right) -\exp \left( -i\omega u\right) \right)  \label{e5b}
\end{equation}

The modes $\varphi _{\omega }(u,v)$ of (\ref{e3}) describe waves incident
from the right as it is clear from the sign of the exponential in the first
term; the second term represents the reflected part which has a rather
complicated behaviour depending on the motion of the mirror. These modes
constitute the {\it in }space and should obviously be unoccupied before
acceleration starts, 
\[
a_{i}\left| 0in\right\rangle =0 
\]
in the language of the following section. Similarly the modes $\bar{\varphi}%
_{\omega }(u,v)$ describe modes travelling to the right (emitted by the
mirror) as can be seen from the exponential of the second term.
Correspondingly the first term is complicated. These modes define the {\it %
out }space and 
\[
\bar{a}_{i}\left| 0out\right\rangle =0 
\]
The state \noindent $\left| 0out\right\rangle $ corresponds to the state
where nothing is produced by the mirror. The two representations are
connected by the Bogolubov transformation to be reviewed in the following
section. The fact that the {\it in }and {\it out }vacua are not identical
lies at the origin of particle production.

\subsection{An illustration}

It is instructive to consider motion at uniform velocity 
\begin{equation}
z=Bt  \label{21a}
\end{equation}

\noindent The function $f$ is given in this case by (\ref{12b}) with $C=0$
(since the condition $f(0)=0$ is valid for the trajectory (\ref{21a})) 
\begin{equation}
f(v)=v%
%TCIMACRO{\dfrac{1-B}{1+B} }
%BeginExpansion
{\displaystyle {1-B \over 1+B}}%
%EndExpansion
\label{e9b}
\end{equation}
\noindent Thus mode (\ref{e3}) reads 
\begin{equation}
\varphi _{\omega }(u,v)=%
%TCIMACRO{\dfrac{i}{2\sqrt{\pi \omega }} }
%BeginExpansion
{\displaystyle {i \over 2\sqrt{\pi \omega }}}%
%EndExpansion
\left( \exp (-i\omega v)-\exp \left( -i\omega u%
%TCIMACRO{\dfrac{1+B}{1-B} }
%BeginExpansion
{\displaystyle {1+B \over 1-B}}%
%EndExpansion
\right) \right)  \label{e8}
\end{equation}

On the other hand in the case of uniform velocity the modes may be obtained
by simply boosting the modes pertaining to the stationary mirror. We denote
the coordinates in the frame of the mirror by $\left( t^{\prime },z^{\prime
}\right) $ and by $\Omega ^{\prime }$ and $K^{\prime }$ the energy and
momentum respectively of a mode in the said frame (of course $\Omega
^{\prime }=\left| K^{\prime }\right| $ but we keep them distinct for the
moment for the sake of clarity). The mode satisfying the boundary condition
at the position of the mirror, taken to be at the origin $z^{\prime }=0$,
reads (modulo a normalization factor) 
\begin{equation}
\exp (-i\Omega ^{\prime }t^{\prime }-iK^{\prime }z^{\prime })-\exp \left(
-i\Omega ^{\prime }t^{\prime }+iK^{\prime }z^{\prime }\right)  \label{e6}
\end{equation}

\noindent where the first term above refers to the incident and the second
to the reflected wave. We express the comoving coordinates in terms of the
lab frame ones via 
\[
z^{\prime }=%
%TCIMACRO{\dfrac{z-Bt}{\sqrt{1-B^{2}}}}
%BeginExpansion
{\displaystyle {z-Bt \over \sqrt{1-B^{2}}}}%
%EndExpansion
\text{ , }t^{\prime }=%
%TCIMACRO{\dfrac{t-Bz}{\sqrt{1-B^{2}}}}
%BeginExpansion
{\displaystyle {t-Bz \over \sqrt{1-B^{2}}}}%
%EndExpansion
\]

\noindent and substitute in (\ref{e6}) to get the mode in the form 
\[
\exp \left( -i%
%TCIMACRO{\dfrac{\Omega ^{\prime }-K^{\prime }B}{\sqrt{1-B^{2}}}}
%BeginExpansion
{\displaystyle {\Omega ^{\prime }-K^{\prime }B \over \sqrt{1-B^{2}}}}%
%EndExpansion
t+i%
%TCIMACRO{\dfrac{-K^{\prime }+\Omega ^{\prime }B}{\sqrt{1-B^{2}}}}
%BeginExpansion
{\displaystyle {-K^{\prime }+\Omega ^{\prime }B \over \sqrt{1-B^{2}}}}%
%EndExpansion
z\right) -\exp \left( -i%
%TCIMACRO{\dfrac{\Omega ^{\prime }+K^{\prime }B}{\sqrt{1-B^{2}}}}
%BeginExpansion
{\displaystyle {\Omega ^{\prime }+K^{\prime }B \over \sqrt{1-B^{2}}}}%
%EndExpansion
t+i%
%TCIMACRO{\dfrac{K^{\prime }+\Omega ^{\prime }B}{\sqrt{1-B^{2}}}}
%BeginExpansion
{\displaystyle {K^{\prime }+\Omega ^{\prime }B \over \sqrt{1-B^{2}}}}%
%EndExpansion
z\right) 
\]

\noindent Using $\Omega ^{\prime }=\left| K^{\prime }\right| $ we rewrite
the above 
\begin{equation}
\exp \left( -i%
%TCIMACRO{\dfrac{\Omega ^{\prime }(1-B)}{\sqrt{1-B^{2}}} }
%BeginExpansion
{\displaystyle {\Omega ^{\prime }(1-B) \over \sqrt{1-B^{2}}}}%
%EndExpansion
(t+z)\right) -\exp \left( -i%
%TCIMACRO{\dfrac{\Omega ^{\prime }(1+B)}{\sqrt{1-B^{2}}} }
%BeginExpansion
{\displaystyle {\Omega ^{\prime }(1+B) \over \sqrt{1-B^{2}}}}%
%EndExpansion
(t-z)\right)  \label{e7}
\end{equation}

\noindent Observe that expression (\ref{e7}) is identical to (\ref{e5}) if
we set 
\begin{equation}
\omega =\frac{\Omega ^{\prime }(1+B)}{\sqrt{1-B^{2}}}=\Omega ^{\prime }\sqrt{%
%TCIMACRO{\dfrac{1+B}{1-B} }
%BeginExpansion
{\displaystyle {1+B \over 1-B}}%
%EndExpansion
}\text{ }  \label{e10}
\end{equation}

\noindent and use (\ref{e9b}). The second term in (\ref{e7}) makes it clear
that the $\omega $\ appearing in (\ref{e5}) is the photon energy in the lab\
frame.

\section{Radiation from a perfect mirror accelerated for a finite time}

\subsection{The Bogolubov transformation}

In this subsection we fix conventions and notation following Birrell and \
Davies op. cit. sec 3.2, write down the general expression for the Bogolubov
amplitude $\beta \left( \omega ,\omega ^{\prime }\right) $ to be evaluated,
and study its asymptotic behaviour for large $\omega ^{\prime }$. The
creation and annihilation operators referring to the two sets $\varphi ,\bar{%
\varphi}$ are connected by 
\begin{equation}
\bar{a}_{i}=\sum_{j}\left( \alpha _{ji}a_{j}+\beta _{ji}^{*}a_{j}^{\dagger
}\right)  \label{e011}
\end{equation}

\noindent Using (\ref{e011}) and its hermitean conjugate we may immediately
verify that the expectation value of the number of excitations of the mode $%
\left( i\right) $ in the $\left| 0in\right\rangle $ vacuum is given by 
\begin{equation}
\left\langle 0in\right| \bar{N}_{i}\left| 0in\right\rangle =\sum_{j}\left|
\beta _{ji}\right| ^{2}  \label{e0011}
\end{equation}

\noindent In our notation the matrix $\beta _{ji}$ is given by the overlap 
\begin{equation}
\beta (\omega ,\omega ^{\prime })=-\left\langle \bar{\varphi}_{\omega
},\varphi _{\omega ^{\prime }}^{*}\right\rangle  \label{e00011}
\end{equation}

\noindent defined as (see Birrell and \ Davies op. cit. equations (2.9),
(3.36)) 
\begin{equation}
\beta (\omega ,\omega ^{\prime })=-i\int_{0}^{\infty }dz\varphi _{\omega
^{\prime }}(z,0)%
%TCIMACRO{\dfrac{\partial }{\partial t} }
%BeginExpansion
{\displaystyle {\partial  \over \partial t}}%
%EndExpansion
\bar{\varphi}_{\omega }(z,0)+i\int_{0}^{\infty }dz\left( 
%TCIMACRO{\dfrac{\partial }{\partial t} }
%BeginExpansion
{\displaystyle {\partial  \over \partial t}}%
%EndExpansion
\varphi _{\omega ^{\prime }}(z,0)\right) \bar{\varphi}_{\omega }(z,0)
\label{e11}
\end{equation}

\noindent The integration in (\ref{e11}) can be over any spacelike
hypersurface. In all cases examined here the mirror is at rest for $t\leq 0$
and the choice $t=0$ for the hypersurface is thus convenient. Notice the
technical simplification that such a trajectory offers. The $in$ modes
evaluated at $t=0$ involve the function $p(-z)$ (i.e. $p$ evaluated at
negative values of the argument), and are thus given by the simple
expression (\ref{e8}) evaluated at $B=0$. The $\bar{\varphi}$ modes are
given by (\ref{e5}) with $f$ depending on the trajectory. Relation (\ref{e11}%
) is rewritten more explicitly in the form ($f$ is a function of $z$ and its
functional form is determined by (\ref{206}) and the initial condition) 
\begin{eqnarray}
\beta (\omega ,\omega ^{\prime }) &=&-i\int_{0}^{\infty }dz%
%TCIMACRO{\dfrac{i}{2\sqrt{\pi \omega ^{\prime }}} }
%BeginExpansion
{\displaystyle {i \over 2\sqrt{\pi \omega ^{\prime }}}}%
%EndExpansion
\left\{ e^{-i\omega ^{\prime }z}-e^{i\omega ^{\prime }z}\right\} \frac{%
\omega }{2\sqrt{\pi \omega }}\left\{ f^{\prime }e^{-i\omega f}-e^{i\omega
z}\right\} +  \label{e100} \\
&&+i\int_{0}^{\infty }dz%
%TCIMACRO{\dfrac{\omega ^{\prime }}{2\sqrt{\pi \omega ^{\prime }}} }
%BeginExpansion
{\displaystyle {\omega ^{\prime } \over 2\sqrt{\pi \omega ^{\prime }}}}%
%EndExpansion
\left\{ e^{-i\omega ^{\prime }z}-e^{i\omega ^{\prime }z}\right\} \frac{i}{2%
\sqrt{\pi \omega }}\left\{ e^{-i\omega f}-e^{i\omega z}\right\}  \nonumber
\end{eqnarray}

\noindent The above expression may be rearranged in the form 
\begin{equation}
\beta (\omega ,\omega ^{\prime })=%
%TCIMACRO{\dfrac{i\omega ^{\prime }}{2\pi \sqrt{\omega \omega ^{\prime }}} }
%BeginExpansion
{\displaystyle {i\omega ^{\prime } \over 2\pi \sqrt{\omega \omega ^{\prime }}}}%
%EndExpansion
\int_{0}^{\infty }dz\sin \left( \omega ^{\prime }z\right) \left\{
e^{-i\omega f}-e^{i\omega z}\right\} +%
%TCIMACRO{\dfrac{i\omega }{2\pi \sqrt{\omega \omega ^{\prime }}} }
%BeginExpansion
{\displaystyle {i\omega  \over 2\pi \sqrt{\omega \omega ^{\prime }}}}%
%EndExpansion
\int_{0}^{\infty }dz\sin \left( \omega ^{\prime }z\right) \left\{ e^{i\omega
z}-f^{\prime }e^{-i\omega f}\right\}  \label{e100b}
\end{equation}

\noindent Notice that both (\ref{e100}) and (\ref{e100b}) are direct
descendants of (\ref{e11}). Notice also the presence of two $f$-independent
(i.e. trajectory independent) terms whose origin is purely kinematic. The
amplitude $\beta (\omega ,\omega ^{\prime })$ will be evaluated for a
specific trajectory in the next subsection.

In this section we confine ourselves to trajectories that in the infinite
past and infinite future describe motion at uniform velocity. In that case
both {\it in }and {\it out } solutions form complete sets of states, and the
standard manipulations with creation and annihilation operators go through.
Completeness is lost (for example) in the case where the mirror follows the
trajectory (studied in the next subsection) $g(t)=-\ln \left( \cosh t\right) 
$ forever, thus asymptotically approaching the null line $v=-\ln 2$. Then
the trajectory does not cut all the characteristics of the wave equation (it
leaves out the $v=${\it const }ones lying above the null line $v=\ln 2$).
The total number of emitted photons is unambiguously given by (\ref{e0011})
after we sum over $i$ or, in our case, integrate over $\omega $ (the
summation over $j$ corresponds to integration over $\omega ^{\prime }):$%
\begin{equation}
N=\int_{0}^{\infty }d\omega \int_{0}^{\infty }d\omega ^{\prime }\left| \beta
(\omega ,\omega ^{\prime })\right| ^{2}e^{-\alpha \left( \omega +\omega
^{\prime }\right) }  \label{80}
\end{equation}

\noindent The small convergence factor $\alpha $ has been introduced in
anticipation of an ultraviolet divergence in (\ref{80}). The fact that
quantities such as the total number of particles produced or the total
emitted energy suffer from such divergences is also encountered in
nonrelativistic calculations in the case of dispersive mirrors (see for
example Barton and Calogeracos (1995)). The corresponding emitted energy is
given by 
\begin{equation}
E=\int_{0}^{\infty }d\omega \omega \int_{0}^{\infty }d\omega ^{\prime
}\left| \beta (\omega ,\omega ^{\prime })\right| ^{2}e^{-\alpha \left(
\omega +\omega ^{\prime }\right) }  \label{80b}
\end{equation}

One obtains the spectrum as a function of $\omega $ after one performs the
first (with respect to $\omega ^{\prime }$) integration in (\ref{80}) above.
The behaviour of the integrand for large $\omega ^{\prime }$ is crucial.
Walker (1985) has asserted that in the case of a trajectory with continuous
velocity the square $\left| \beta (\omega ,\omega ^{\prime })\right| ^{2}$
goes as $\left( \omega ^{\prime }\right) ^{-5}$. We present a proof based on
the technique of Lighthill's (1958), Chapter 4. We introduce the two
integrals 
\begin{equation}
J_{1}\left( \omega ^{\prime }\right) \equiv \int_{0}^{\infty }dze^{-i\omega
^{\prime }z}\left\{ e^{-i\omega f(z)}-e^{i\omega z}\right\}  \label{800}
\end{equation}
\begin{equation}
J_{2}\left( \omega ^{\prime }\right) \equiv \int_{0}^{\infty }dze^{-i\omega
^{\prime }z}\left\{ f^{\prime }(z)e^{-i\omega f(z)}-e^{i\omega z}\right\}
\label{800b}
\end{equation}
Then the first and second integrals appearing in (\ref{e100b}) are
proportional to 
\begin{equation}
J_{1}\left( -\omega ^{\prime }\right) -J_{1}\left( \omega ^{\prime }\right)
\label{801}
\end{equation}
and 
\begin{equation}
J_{2}\left( -\omega ^{\prime }\right) -J_{2}\left( \omega ^{\prime }\right)
\label{801b}
\end{equation}
respectively. We now examine the asymptotic estimates for large $\omega
^{\prime }$ of $J_{1}\left( \omega ^{\prime }\right) $ and $J_{2}\left(
\omega ^{\prime }\right) $.

We define the functions 
\begin{equation}
\Psi _{A}\left( z\right) \equiv \exp \left( -i\omega f_{in}\left( z\right)
\right) -\exp \left( i\omega z\right) ,\Psi _{B}\left( z\right) \equiv \exp
\left( -i\omega f_{acc}\left( z\right) \right) -\exp \left( i\omega z\right)
,\Psi _{C}\left( z\right) \equiv \exp \left( -i\omega f_{0}\left( z\right)
\right) -\exp \left( i\omega z\right)  \label{802}
\end{equation}
where the functions $f_{in},f_{acc},f_{0}$ are defined by (\ref{12b}) for $%
B=0,C=0$, (\ref{208b}), and (\ref{12b}) for velocity equal to $\beta _{P}$
respectively. In the latter case $C$ is fixed by the requirement that the
velocity be continuous at {\it P} (see(\ref{40b})). To conform with
Lighthill's notation we define $\omega \equiv 2\pi y$. Then $J_{1}\left(
\omega ^{\prime }\right) $ is the Fourier transform $G(y)$ of 
\begin{equation}
\Psi (z)\equiv \Psi _{A}\left( z\right) H(-z)+\Psi _{B}\left( z\right)
H(z)+\left[ \Psi _{C}\left( z\right) -\Psi _{B}\left( z\right) -\Psi
_{A}\left( z\right) \right] H(z-r)  \label{803}
\end{equation}
where $H$ is the Heaviside unit function. The singular points of $\Psi (z)$
are $z=0$ and $z=r$. Clearly the three functions $\Psi _{A}\left( z\right)
,\Psi _{B}\left( z\right) ,\Psi _{C}\left( z\right) $ correspond to the
three stages of a stationary mirror, accelerating mirror, and mirror moving
at uniform velocity respectively. Observe that the general connection
between the velocity $\beta $ and the function $f^{\prime }$ is given by (%
\ref{206}) and that the velocity is everywhere continuous along the
trajectory considered whereas the acceleration has different (but finite)
left and right derivatives at each of the two singular points. Thus $\beta
^{\prime \prime }$ (the {\it second} time derivative of the velocity) has a
finite jump at each of the two singular points and hence is absolutely
integrable everywhere. To treat the first singular point let us define the
function 
\begin{eqnarray}
F_{1}(z) &=&\left[ \Psi _{A}\left( 0\right) +z\Psi _{A}^{\prime }\left(
0\right) +\frac{z^{2}}{2!}\Psi _{A}^{\prime \prime }\left( 0\right) +\frac{%
z^{3}}{3!}\Psi _{A}^{\prime \prime \prime }\left( 0\right) \right] H(-z)+
\label{806} \\
&&  \nonumber \\
&&+\left[ \Psi _{B}\left( 0\right) +z\Psi _{B}^{\prime }\left( 0\right) +%
\frac{z^{2}}{2!}\Psi _{B}^{\prime \prime }\left( 0\right) +\frac{z^{3}}{3!}%
\Psi _{A}^{\prime \prime \prime }\left( 0\right) \right] H(z)  \nonumber
\end{eqnarray}
According to what was said above the quantity $\Psi _{A}^{\prime \prime
\prime }\left( 0\right) -\Psi _{B}^{\prime \prime \prime }\left( 0\right) $
is non-vanishing but finite. Hence the function $\Psi (z)-F_{1}(z)$ has
absolutely integrable third derivative at $z=0$. We treat the second
singular point $z=r$ by a similar method, defining the function 
\begin{eqnarray}
F_{2}(z) &=&\left[ \Psi _{B}\left( r\right) +(r-z)\Psi _{B}^{\prime }\left(
r\right) +\frac{(r-z)^{2}}{2!}\Psi _{B}^{\prime \prime }\left( r\right) +%
\frac{(r-z)^{3}}{3!}\Psi _{B}^{\prime \prime \prime }\left( r\right) \right]
H(r-z)+  \label{807} \\
&&  \nonumber \\
&&+\left[ \Psi _{C}\left( r\right) +(z-r)\Psi _{C}^{\prime }\left( r\right) +%
\frac{(z-r)^{2}}{2!}\Psi _{C}^{\prime \prime }\left( 0\right) +\frac{%
(z-r)^{3}}{3!}\Psi _{C}^{\prime \prime \prime }\left( 0\right) \right] H(z-r)
\nonumber
\end{eqnarray}
As before the quantity $\Psi _{C}^{\prime \prime \prime }\left( r\right)
-\Psi _{B}^{\prime \prime \prime }\left( r\right) $ is non-vanishing but
finite. Finally observe that the function $\Psi ^{\prime \prime \prime }(z)$
is well-behaved at infinity in the sense of Lighthill's definition 20, p.
49. Then according to Lighthill's theorem 19, p. 52 the Fourier transform $%
G(y)$ of $\Psi (z)$ takes the form 
\begin{equation}
G(y)=G_{1}(y)+G_{2}(y)+o\left( \left| y\right| ^{-3}\right)  \label{808}
\end{equation}
where $G_{1}(y),G_{2}(y)$ are the Fourier transforms of $F_{1}(z),F_{2}(z)$.
According to Lighthill's table, p. 43 the functions $G_{1}(y),G_{2}(y)$ go
asymptotically as $y^{-4}$. Hence overall $G(y)\approx o\left( \left|
y\right| ^{-3}\right) $. Returning to (\ref{800}) we observe that $%
J_{1}\left( -\omega ^{\prime }\right) $ goes as $\left( \omega ^{\prime
}\right) ^{-3}$ and thus so does the difference (\ref{801}). Regarding the
integral (\ref{800b}) we may repeat the above process almost verbatim. The
one difference is that this integral involves $f^{\prime }$, thus in the
previous argument one has to substitute ''second derivative'' for ''third
derivative'' and accordingly keep terms of second order rather than third in
(\ref{806}), (\ref{807}). Hence the second integral goes as $\left( \omega
^{\prime }\right) ^{-2}$. Noticing the prefactors in (\ref{e100b}) and
especially the fact that the first integral is multiplied by one power of $%
\omega ^{\prime }$ we deduce that overall $\left| \beta (\omega ,\omega
^{\prime })\right| ^{2}$ indeed goes as $\left( \omega ^{\prime }\right)
^{-5}.$ It is shown in section 3 of II that the above arguments cannot be
applied in the case of a mirror accelerating forever and that in that case $%
\left| \beta (\omega ,\omega ^{\prime })\right| ^{2}$ behaves as $1/\omega
^{\prime }.$

\subsection{The Bogolubov amplitudes for a mirror eventually reverting to
uniform velocity}

In this section we consider a trajectory defined as follows. The mirror is
stationary for $t<0$, follows the trajectory

\begin{equation}
z=g(t)=-\ln (\cosh t)  \label{e20a}
\end{equation}

\noindent described in Appendix A till some spacetime point {\it P}, and
then continues at uniform velocity $\beta _{P}$ (figure 2). The velocity is
continuous at $O$ and at {\it P} whereas of course the acceleration $%
a=d\beta /dt$ is not. The trajectory equation (\ref{e20a}) is a special case
of $z=-\frac{1}{\kappa }\ln (\cosh (\kappa t))$. The energy scale is fixed
in (\ref{e20a}) by setting $\kappa =1$. Our interest in the trajectory (\ref
{e20a}) stems from the fact that if (and only if) the acceleration is
allowed to go on forever then the mirror gives rise to a thermal spectrum of
radiation. For details see II and references therein. Notice that for large $%
t$ (\ref{e20a}) has the asymptotic form 
\begin{equation}
g(t)\approx -t-e^{-2t}+\ln 2  \label{690}
\end{equation}

\noindent \noindent The advantages of an asymptotically inertial trajectory
have been mentioned earlier on.

Let $r$ be the $z$ intercept of the null line passing through {\it P} with
slope -1. As can be seen from the equation of the trajectory $v=p(u)$ with $%
p(u)$ given by (\ref{a3}) the possible values of $r$ range from $0$ to $\ln
2 $ (the latter asymptotically when $u\rightarrow \infty $). The quantities $%
r$ and $\beta _{P}$ are connected via (\ref{a2}) 
\[
\beta _{P}=1-e^{r} 
\]

When it comes to the integrations appearing in the amplitude (\ref{e100}),
we use for $f(z)$ the expression $f_{acc}(z)$ given by (\ref{208b}) in the
range $0\leq z\leq r$ and $f_{0}(z)$ given by (\ref{12b}) in the range $%
r\leq z\leq \infty $. Some of the integrals involved are conveniently
expressed in terms of the function $\zeta $ and its complex conjugate $\zeta
^{*}$defined in Heitler (1954), pages 66-71: 
\begin{equation}
\zeta (x)\equiv -i\int_{0}^{\infty }e^{i\kappa x}d\kappa =P\frac{1}{x}-i\pi
\delta (x)  \label{delta}
\end{equation}

\noindent The constant $C$ in (\ref{12b}) is determined by continuity of $f$
at the point {\it P }

\[
f_{acc}(z)=f_{0}(z) 
\]

\noindent or

\begin{equation}
C+\frac{1-\beta _{P}}{1+\beta _{P}}r=-\ln \left( 2-e^{r}\right)  \label{40b}
\end{equation}

\noindent where we used (\ref{a4}) for the right hand side of (\ref{40b}).

Amplitude (\ref{e100b}) is split to three contributions

\begin{equation}
\beta (\omega ,\omega ^{\prime })\equiv \beta _{I}(\omega ,\omega ^{\prime
})+\beta _{II}(\omega ,\omega ^{\prime })+\beta _{III}(\omega ,\omega
^{\prime })  \label{e50}
\end{equation}

\noindent originating as follows. The $\beta _{III}(\omega ,\omega ^{\prime
})$ stands for the contribution of the two terms in (\ref{e100b}) that are $%
f $-independent, is always there for a mirror starting from rest (or
initially moves at uniform velocity), and does not depend on the subsequent
form of the trajectory. The $\beta _{I}(\omega ,\omega ^{\prime })$
contribution results from the accelerating part of the trajectory and
corresponds to the $0<z<r$ integration range in (\ref{e100b}). Its
evaluation will prove mathematically somewhat involved. The $\beta
_{II}(\omega ,\omega ^{\prime }) $ amplitude given results from the $z>r$
part of the integration range in (\ref{e100b}), is readily evaluated, and is
specific to a mirror that reverts to the state of uniform motion. Thus

\begin{equation}
\beta _{I}(\omega ,\omega ^{\prime })=-%
%TCIMACRO{\dfrac{i}{2\pi \sqrt{\omega ^{\prime }\omega }} }
%BeginExpansion
{\displaystyle {i \over 2\pi \sqrt{\omega ^{\prime }\omega }}}%
%EndExpansion
\int_{0}^{r}dz\sin (\omega ^{\prime }z)\left\{ \omega f_{acc}^{\prime
}(z)-\omega ^{\prime }\right\} e^{-i\omega f_{acc}(z)}  \label{e20}
\end{equation}

\[
\beta _{II}(\omega ,\omega ^{\prime })= 
\]

\[
=\frac{1}{4\pi i\sqrt{\omega \omega ^{\prime }}}\left( \omega 
%TCIMACRO{\dfrac{1-\beta _{P}}{1+\beta _{P}}}
%BeginExpansion
{\displaystyle {1-\beta _{P} \over 1+\beta _{P}}}%
%EndExpansion
-\omega ^{\prime }\right) \exp \left[ i\left( \omega 
%TCIMACRO{\dfrac{1-\beta _{P}}{1+\beta _{P}}}
%BeginExpansion
{\displaystyle {1-\beta _{P} \over 1+\beta _{P}}}%
%EndExpansion
+\omega ^{\prime }\right) r\right] \zeta \left( \omega 
%TCIMACRO{\dfrac{1-\beta _{P}}{1+\beta _{P}}}
%BeginExpansion
{\displaystyle {1-\beta _{P} \over 1+\beta _{P}}}%
%EndExpansion
+\omega ^{\prime }\right) e^{-i\omega C}- 
\]

\begin{equation}
-\frac{1}{4\pi i\sqrt{\omega \omega ^{\prime }}}\exp \left[ i\left( -\omega
^{\prime }+\omega 
%TCIMACRO{\dfrac{1-\beta _{P}}{1+\beta _{P}} }
%BeginExpansion
{\displaystyle {1-\beta _{P} \over 1+\beta _{P}}}%
%EndExpansion
\right) r\right] e^{-i\omega C}  \label{e102}
\end{equation}

\begin{equation}
\beta _{III}(\omega ,\omega ^{\prime })=\frac{1}{4\pi i\sqrt{\omega \omega
^{\prime }}}-\frac{1}{4\pi i\sqrt{\omega \omega ^{\prime }}}\left( \omega
-\omega ^{\prime }\right) \zeta \left( \omega +\omega ^{\prime }\right)
\label{e103}
\end{equation}

\noindent Notice that the arguments of the $\zeta $ functions in (\ref{e102}%
) and (\ref{e103}) never vanish. Thus as far as the evaluation of $\beta
(\omega ,\omega ^{\prime })$ is concerned we observe that it is only the
first term in (\ref{delta}) that is operative, and that $\beta (\omega
,\omega ^{\prime })$ does not have any $\delta $-type singularities. However
the forms (\ref{e102}), (\ref{e103}) are useful because we can obtain the
other Bogolubov amplitude $\alpha (\omega ,\omega ^{\prime })$ via the
substitution $\omega \rightarrow -\omega $ in accordance with (\ref{e107})
below. The $\alpha (\omega ,\omega ^{\prime })$ is of course expected to
have $\delta $-type singularities, and in fact reduces to just $\delta
(\omega -\omega ^{\prime })$ in the trivial case when the {\it in }and {\it %
out }modes coincide.

We turn our attention to $\beta _{I}(\omega ,\omega ^{\prime })$. We
integrate (\ref{e20}) by parts to get

\begin{equation}
\beta _{I}(\omega ,\omega ^{\prime })=\frac{1}{2\pi \sqrt{\omega \omega
^{\prime }}}\sin \left( \omega ^{\prime }r\right) e^{-i\omega f_{acc}(r)}-%
\frac{1}{2\pi }\sqrt{\frac{\omega ^{\prime }}{\omega }}\int_{0}^{r}dze^{-i%
\omega f_{acc}(z)-i\omega ^{\prime }z}  \label{e20b}
\end{equation}

\noindent where the first term is the upper end-point contribution; the
lower end-point contribution vanishes. We rewrite (\ref{e20b}) using (\ref
{208b}) for $f_{acc}$%
\begin{equation}
\beta _{I}(\omega ,\omega ^{\prime })=\frac{1}{2\pi \sqrt{\omega \omega
^{\prime }}}\sin \left( \omega ^{\prime }r\right) \left( 2-e^{r}\right)
^{i\omega }-%
%TCIMACRO{\dfrac{1}{2\pi } }
%BeginExpansion
{\displaystyle {1 \over 2\pi }}%
%EndExpansion
\sqrt{\frac{\omega ^{\prime }}{\omega }}\int_{0}^{r}dz\left( 2-e^{z}\right)
^{i\omega }e^{-i\omega ^{\prime }z}  \label{e22}
\end{equation}

\noindent We make the change of variable

\begin{equation}
z=\ln 2-\rho  \label{e22a}
\end{equation}

\noindent and rewrite (\ref{e22}) in the form 
\begin{equation}
\beta _{I}(\omega ,\omega ^{\prime })=\frac{1}{2\pi \sqrt{\omega \omega
^{\prime }}}\sin \left( \omega ^{\prime }r\right) \left( 2-e^{r}\right)
^{i\omega }-%
%TCIMACRO{\dfrac{2^{i\left( \omega -\omega ^{\prime }\right) }}{2\pi } }
%BeginExpansion
{\displaystyle {2^{i\left( \omega -\omega ^{\prime }\right) } \over 2\pi }}%
%EndExpansion
\sqrt{\frac{\omega ^{\prime }}{\omega }}\int_{\ln 2-r}^{\ln 2}d\rho \left(
1-e^{-\rho }\right) ^{i\omega }e^{i\omega ^{\prime }\rho }  \label{e23}
\end{equation}

\noindent \noindent We isolate the integral occurring in (\ref{e23}) and
rewrite it with arbitrary limits 
\begin{equation}
I\equiv \int_{l_{1}}^{l_{2}}d\rho \left( 1-e^{-\rho }\right) ^{i\omega
}e^{i\omega ^{\prime }\rho }  \label{e24}
\end{equation}

\noindent \noindent We use the binomial expansion in (\ref{e24}) and
integrate term by term to get 
\begin{equation}
I=\left[ -e^{i\omega ^{\prime }x}\sum_{n=0}^{\infty }\frac{\left( -i\omega
\right) _{n}}{n!}\frac{\left( e^{-x}\right) ^{n}}{\left( n-i\omega ^{\prime
}\right) }\right] _{x=l_{1}}^{x=l_{2}}  \label{b10}
\end{equation}

\noindent where $\left( \cdot \right) _{n}$ is the Pochhammer symbol. We
write 
\begin{equation}
\frac{1}{\left( n-i\omega ^{\prime }\right) }=\frac{\left( -i\omega ^{\prime
}\right) _{n}}{\left( -i\omega ^{\prime }+1\right) _{n}}  \label{b10a}
\end{equation}

\noindent and then the sum in (\ref{b10}) is identified as the $F$
hypergeometric function: 
\begin{equation}
I=\left[ \frac{e^{i\omega ^{\prime }x}}{i\omega ^{\prime }}F\left( -i\omega
,-i\omega ^{\prime };-i\omega ^{\prime }+1;e^{-x}\right) \right]
_{x=l_{1}}^{x=l_{2}}  \label{b11}
\end{equation}

\noindent Notice that according to item 15.1.1 (b) of Abramowitz and Stegun
(1972) the series expansion of the $F(a,b,c,z)$ hypergeometric converges
absolutely if $%
%TCIMACRO{\func{Re}}
%BeginExpansion
\mathop{\rm Re}%
%EndExpansion
(c-a-b)=1>0$ and this inequality does hold for the hypergeometric occurring
in (\ref{b11}). We reinstate the limits $l_{1}=\ln 2-r,l=\ln 2$ appearing in
(\ref{e23}) and obtain for the $\beta _{I}(\omega ,\omega ^{\prime })$
amplitude the expression

\noindent 
\begin{equation}
\beta _{I}(\omega ,\omega ^{\prime })=\frac{1}{2\pi \sqrt{\omega \omega
^{\prime }}}\sin \left( \omega ^{\prime }r\right) \left( 2-e^{r}\right)
^{i\omega }-  \label{e25}
\end{equation}

\[
-\frac{1}{2\pi i}\frac{2^{i\omega }}{\sqrt{\omega \omega ^{\prime }}}%
\{F\left( -i\omega ,-i\omega ^{\prime };-i\omega ^{\prime }+1;\frac{1}{2}%
\right) + 
\]

\[
+F\left( -i\omega ,-i\omega ^{\prime };-i\omega ^{\prime }+1;\frac{e^{r}}{2}%
\right) e^{-i\omega ^{\prime }r}\} 
\]

\noindent It is interesting to observe that the value $r=\ln 2$
corresponding to the asymptote falls exactly on the radius of convergence of
the series as one can see from the argument of the second hypergeometric
appearing above. (Notice that according to (\ref{a2}) $e^{r}=1-\beta _{P},$
recall that $\beta _{P}$ is the velocity at {\it P,} and that asymptotically 
$\beta \rightarrow -1.$)

An alternative way to evaluate (\ref{e24}) is to rewrite the integral
occurring in (\ref{e23})

\begin{equation}
I=\int_{\ln 2-r}^{\ln 2}d\rho \left( 1-e^{-\rho }\right) ^{i\omega
}e^{i\omega ^{\prime }\rho -\alpha \rho }  \label{e111}
\end{equation}

\noindent introducing a small positive constant $\alpha $ to ensure
convergence. We change variable to 
\[
t=e^{-\rho } 
\]

\noindent Then (\ref{e111}) reads 
\[
I=\int_{1/2}^{e^{r}/2}dtt^{-1-i\omega ^{\prime }+\alpha }\left( 1-t\right)
^{i\omega }=\int_{0}^{e^{r}/2}dtt^{-1-i\omega ^{\prime }+\alpha }\left(
1-t\right) ^{i\omega }-\int_{0}^{1/2}dtt^{-1-i\omega ^{\prime }+\alpha
}\left( 1-t\right) ^{i\omega } 
\]

\noindent If we make use of the integral representation 15.3.1 of Abramowitz
and Stegun op. cit. for the hypergeometric function and take the $\alpha
\rightarrow 0$ limit we get for $I$ the result (\ref{b11}).

Relation (\ref{e25}) provides the final result for $\beta _{I}(\omega
,\omega ^{\prime })$. However for a numerical evaluation (via MAPLE\ for
example) of the second hypergeometric 
\begin{equation}
F\left( -i\omega ,-i\omega ^{\prime };-i\omega ^{\prime }+1;\frac{e^{r}}{2}%
\right)  \label{e25d}
\end{equation}

\noindent occurring in (\ref{e25}) and if the point {\it P} is near the
asymptote (i.e. $e^{r}/2\simeq 1$) one can accelerate the convergence
through the use of the linear transformation formula 15.3.6 of Abramowitz \&
Stegun op. cit. 
\begin{eqnarray}
F\left( a,b;c;z\right) &=&F\left( a,b;a+b-c+1;1-z\right) \frac{\Gamma \left(
c\right) \Gamma \left( c-a-b\right) }{\Gamma \left( c-a\right) \Gamma \left(
c-b\right) }+  \label{e25e} \\
&&F\left( c-a,c-b;c-a-b+1;1-z\right) \left( 1-z\right) ^{c-a-b}\frac{\Gamma
\left( c\right) \Gamma \left( a+b-c\right) }{\Gamma (a)\Gamma (b)}  \nonumber
\end{eqnarray}
Using our $a,b,c$ the first hypergeometric in (\ref{e25e}) reads $F\left(
-i\omega ,-i\omega ^{\prime };-i\omega ;1-e^{r}/2\right) $. We expand it in
the standard hypergeometric series and for this particular set of parameters
the series reduces to the simple binomial expansion and get 
\[
F\left( -i\omega ,-i\omega ^{\prime };-i\omega ;1-e^{r}/2\right) =\left( 
\frac{e^{r}}{2}\right) ^{\omega ^{\prime }} 
\]

\noindent We may thus rewrite the amplitude (\ref{e25}) 
\begin{eqnarray}
&&\beta _{I}(\omega ,\omega ^{\prime })  \label{e26b} \\
&=&\frac{1}{2\pi \sqrt{\omega \omega ^{\prime }}}\sin \left( \omega ^{\prime
}r\right) \left( 2-e^{r}\right) ^{i\omega }-  \nonumber \\
&&-\frac{1}{2\pi i}\frac{2^{i\omega }}{\sqrt{\omega \omega ^{\prime }}}%
\{F\left( -i\omega ,-i\omega ^{\prime };-i\omega ^{\prime }+1;\frac{1}{2}%
\right) +  \nonumber
\end{eqnarray}

\[
+\frac{\Gamma (-i\omega ^{\prime }+1)\Gamma (i\omega +1)}{\Gamma (1-i\omega
^{\prime }+i\omega )}\left( \frac{e^{r}}{2}\right) ^{\omega ^{\prime
}}e^{-i\omega ^{\prime }r}+ 
\]

\[
+F\left( -i\omega ^{\prime }+1+i\omega ,1;i\omega +2;1-\frac{e^{r}}{2}%
\right) \frac{\Gamma (-i\omega ^{\prime }+1)\Gamma (-i\omega -1)}{\Gamma
(-i\omega ^{\prime })\Gamma \left( -i\omega \right) }\left( 1-\frac{e^{r}}{2}%
\right) ^{1+i\omega }e^{-i\omega ^{\prime }r}\} 
\]

\noindent To summarize, the advantage of the above form over (\ref{e25})
lies in the fact that for a point $P$ near the asymptotic line the arguments
of the hypergeometrics are far from unity (the radius of convergence). In
conclusion the sum of (\ref{e102}), (\ref{e103}) and (\ref{e26b}) is the
exact answer for the $\beta \left( \omega ,\omega ^{\prime }\right) $
amplitude for the trajectory in question.

\subsection{On the Bogolubov coefficients}

The Bogolubov $\alpha $ coefficients appearing in (\ref{e011}) are given by
the analog of (\ref{e00011}) 
\begin{equation}
\alpha (\omega ,\omega ^{\prime })=\left\langle \bar{\varphi}_{\omega
},\varphi _{\omega ^{\prime }}\right\rangle  \label{90}
\end{equation}

\noindent or explicitly by 
\begin{equation}
\alpha (\omega ,\omega ^{\prime })=i\int_{0}^{\infty }dz\varphi _{\omega
^{\prime }}(z,0)%
%TCIMACRO{\dfrac{\partial }{\partial t} }
%BeginExpansion
{\displaystyle {\partial  \over \partial t}}%
%EndExpansion
\bar{\varphi}_{\omega }^{*}(z,0)-i\int_{0}^{\infty }dz\left( 
%TCIMACRO{\dfrac{\partial }{\partial t} }
%BeginExpansion
{\displaystyle {\partial  \over \partial t}}%
%EndExpansion
\varphi _{\omega ^{\prime }}(z,0)\right) \bar{\varphi}_{\omega }^{*}(z,0)
\label{91}
\end{equation}

\noindent Recall also the unitarity condition ((3.39) of Birrell and \
Davies op. cit.) 
\begin{equation}
\int_{0}^{\infty }d\tilde{\omega}\left( \alpha \left( \tilde{\omega},\omega
_{1}\right) \alpha ^{*}\left( \tilde{\omega},\omega _{2}\right) -\beta
\left( \tilde{\omega},\omega _{1}\right) \beta ^{*}\left( \tilde{\omega}%
,\omega _{2}\right) \right) =\delta \left( \omega _{1}-\omega _{2}\right)
\label{e92a}
\end{equation}

\noindent and its partner 
\begin{equation}
\int_{0}^{\infty }d\widetilde{\omega }\left( \alpha \left( \omega _{1},%
\widetilde{\omega }\right) \alpha ^{*}\left( \omega _{2},\widetilde{\omega }%
\right) -\beta \left( \omega _{1},\widetilde{\omega }\right) \beta
^{*}\left( \omega _{2},\widetilde{\omega }\right) \right) =\delta \left(
\omega _{1}-\omega _{2}\right)  \label{92}
\end{equation}
Relations (\ref{e92a}), (\ref{92}) above are direct consequences of the fact
that the sets $\bar{\varphi}_{\omega }$ and $\varphi _{\omega ^{\prime }}$
respectively are orthonormal and complete. They also guarantee that the
operators $a_{j},a_{j}^{\dagger }$ and $\bar{a}_{i},\bar{a}_{i}^{\dagger }$
obey the standard equal time commutation relations that creation and
annihilation operators do.

We find it convenient to isolate the square roots in (\ref{e100b}) and
introduce quantities $A(\omega ,\omega ^{\prime }),B(\omega ,\omega ^{\prime
})$ that are analytic functions of the frequencies (without the branch cuts
attendant to square roots) via 
\begin{equation}
\alpha (\omega ,\omega ^{\prime })=\frac{A(\omega ,\omega ^{\prime })}{\sqrt{%
\omega \omega ^{\prime }}},\beta (\omega ,\omega ^{\prime })=\frac{B(\omega
,\omega ^{\prime })}{\sqrt{\omega \omega ^{\prime }}}  \label{e106}
\end{equation}

The quantity $B(\omega ,\omega ^{\prime })$ is read off (\ref{e100}) (and $%
A(\omega ,\omega ^{\prime })$ from the corresponding expression for $\alpha
(\omega ,\omega ^{\prime })$). From the definitions (\ref{e00011}) and (\ref
{90}) of the Bogolubov coefficients, the explicit form (\ref{e100}) of the
overlap integral and expression (\ref{e5b}) for the field derivative one can
immediately deduce that 
\begin{equation}
B^{*}(\omega ,\omega ^{\prime })=A(-\omega ,\omega ^{\prime }),A^{*}(\omega
,\omega ^{\prime })=B(-\omega ,\omega ^{\prime })  \label{e107}
\end{equation}

\noindent Observe that identities (\ref{e92a}), (\ref{92}) are a result of
the completeness of the basis {\it out }and {\it in }wavefunctions
respectively.

\section{Radiation from a dispersive mirror}

We shall repeat the construction of section II in the case of a dispersive
mirror. We adopt the model of Barton and Calogeracos (1995) and Calogeracos
and Barton (1995). The model is described by the covariant Action 
\begin{equation}
I=\int dt\left\{ \int_{-\infty }^{\xi -}+\int_{\xi +}^{+\infty }\right\}
dz\left\{ \frac{1}{2}g^{\mu \nu }\partial _{\mu }\partial _{\nu }\Phi
\right\} -\gamma \int d\tau \Phi ^{2}\left( \xi \right) -M\int d\tau
\label{52a}
\end{equation}

\noindent The constant $\gamma $ controls the dispersivity of the mirror.
The mirror follows a prescribed trajectory. The statement following equation
(\ref{e1}) remains true in the case of a dispersive mirror. Any linear
combination of functions depending on $u$ or $v$ separately is a solution of
(\ref{e1}). Outgoing modes will be denoted by an overbar (as in section 2),
incoming modes without one. Indices (+) and ($-$) correspond to the space
argument of $\Phi $ being to the right and to the left of the mirror
respectively. Notice that eventually the expressions for the modes will be
substituted in (\ref{e11}). Since we evaluate the overlap integral at $t=0$,
the {\it in }modes to be used coincide with the usual modes for a stationary
mirror as exhibited in Barton and Calogeracos (1995). The effects of the
motion are manifested in the complicated appearance of the {\it out }modes.
This is of course exactly the same state of affairs that we have already
encountered in the case of a perfect mirror. The general expression for the 
{\it in }modes is given for the sake of completeness.

Incident modes (figures 3 and 4) can now come from either direction. From the right
they read 
\begin{equation}
\Phi _{(+)R\omega }\left( z,t\right) =\frac{i}{2\sqrt{\pi \omega }}\left\{
e^{-i\omega v}+R_{R}(\omega ,p(u))e^{-i\omega p(u)}\right\}  \label{52}
\end{equation}
\begin{equation}
\Phi _{(-)R\omega }\left( z,t\right) =\frac{i}{2\sqrt{\pi \omega }}%
T_{R}(\omega ,v)e^{-i\omega v}  \label{53}
\end{equation}

\noindent Notice that each term in the above equations depends either on $u$
or $v$ only, hence solves (\ref{e1}). Similarly a left-incident wave is
written 
\begin{equation}
\Phi _{(+)L\omega }\left( z,t\right) =\frac{i}{2\sqrt{\pi \omega }}%
T_{L}(\omega ,u)e^{-i\omega u}  \label{54}
\end{equation}
\begin{equation}
\Phi _{(-)L\omega }\left( z,t\right) =\frac{i}{2\sqrt{\pi \omega }}\left\{
e^{-i\omega u}+R_{L}(\omega ,f(v))e^{-i\omega f(v)}\right\}  \label{55}
\end{equation}

The above modes constitute the {\it in }states. The {\it out }modes
correspond to waves transmitted to either right (figure 5) or left (figure 6). Left transmitted
ones read 
\begin{equation}
\bar{\Phi}_{(+)R\omega }\left( z,t\right) =\frac{i}{2\sqrt{\pi \omega }}%
\left\{ e^{-i\omega f(v)}+\bar{R}_{R}(\omega ,u)e^{-i\omega u}\right\}
\label{70}
\end{equation}

\begin{equation}
\bar{\Phi}_{(-)R\omega }\left( z,t\right) =\frac{i}{2\sqrt{\pi \omega }}\bar{%
T}_{R}(\omega ,f(v))e^{-i\omega v}  \label{71}
\end{equation}
The right transmitted ones have the form 
\begin{equation}
\bar{\Phi}_{(+)L\omega }\left( z,t\right) =\frac{i}{2\sqrt{\pi \omega }}\bar{%
T}_{L}(p(u))e^{-i\omega p(u)}  \label{72}
\end{equation}
\begin{equation}
\bar{\Phi}_{(-)L\omega }\left( z,t\right) =\frac{i}{2\sqrt{\pi \omega }}%
\left\{ e^{-i\omega p(u)}+\bar{R}_{L}(\omega ,v)e^{-i\omega v}\right\}
\label{73}
\end{equation}
Continuity of the modes on the trajectory is guaranteed by the trajectory
equation $u=f(v)$ (\ref{e005}), its inverse $v=p(u)$ (\ref{e05}), and by 
\begin{equation}
T=1+R  \label{56}
\end{equation}

\noindent which is true for all the above modes due to the continuity of $%
\Phi $ on the trajectory (indices, overbars and arguments suppressed). The $%
T $s and the $R$s for a moving mirror are nontrivial functions of spacetime
and their dependence on $u$ or $v$ as the case may be will be determined
below. We quote the $R$ and $T$ for a mirror at rest as given in \cite{BC}: 
\begin{equation}
R(\omega )=-\frac{i\gamma }{\omega +i\gamma },T(\omega )=\frac{\omega }{%
\omega +i\gamma }  \label{rt}
\end{equation}

Concerning the mode expressions notice that the functional form of the left
incident modes (\ref{54}), (\ref{55}) (figure 3) is identical to that of
the right incident modes (\ref{52}), (\ref{53}) (figure 4) pertaining to a
mirror moving along the trajectory $-g(t)$ (reflected with respect to the
origin). This is manifested by the fact that (\ref{54}) and (\ref{55}) are
obtained from (\ref{53}) and (\ref{52}) respectively under the
correspondence $R\rightleftharpoons L,$ $v\rightleftharpoons u,$ $%
f\rightleftharpoons p$. In this respect recall the remark after (\ref{e005b}%
). An analogous statement holds about the transmitted modes (\ref{70}) to (%
\ref{73}).

It should be noticed that for a right incident wave the relation (\ref{e10})
between the photon energy $\omega $ in the lab frame and the photon energy $%
\Omega ^{\prime }$ in the mirror's rest frame holds true. For a left
incident wave the analog of (\ref{e10}) reads 
\begin{equation}
\omega =\Omega ^{\prime }%
%TCIMACRO{\dfrac{1-B}{\sqrt{1-B^{2}}} }
%BeginExpansion
{\displaystyle {1-B \over \sqrt{1-B^{2}}}}%
%EndExpansion
\label{59c}
\end{equation}

The discontinuity of $\partial _{\mu }\Phi $ across the trajectory is given
in Calogeracos and Barton (1995) (cf their equation\ (II.2) or more
explicitly \ (III.5), after the latter has been transformed to the lab
frame): 
\begin{equation}
disc\left[ \beta 
%TCIMACRO{\dfrac{\partial \Phi }{\partial t} }
%BeginExpansion
{\displaystyle {\partial \Phi  \over \partial t}}%
%EndExpansion
+%
%TCIMACRO{\dfrac{\partial \Phi }{\partial x} }
%BeginExpansion
{\displaystyle {\partial \Phi  \over \partial x}}%
%EndExpansion
\right] _{(-)}^{(+)}=2\gamma \Phi \sqrt{1-\beta ^{2}}  \label{57}
\end{equation}

\noindent where $\beta =dg(t)/dt$ is the velocity (and $z=g(t)$ the
trajectory equation). The quantity $\Phi $ in the right hand side of the
above relation is evaluated on the trajectory.

We substitute the modes in the boundary condition (\ref{57}), recall that
the derivatives in (\ref{57}) also act on $R$ and $T$ and use (\ref{56}) to
get in a straightforward manner a set of four (decoupled) differential
equations, each for a set of modes.

(i) Modes transmitted to the left (equations (\ref{70}), (\ref{71})): 
\begin{equation}
%TCIMACRO{\dfrac{2}{i\omega } }
%BeginExpansion
{\displaystyle {2 \over i\omega }}%
%EndExpansion
%TCIMACRO{\dfrac{d\bar{R}_{R}(u)}{du} }
%BeginExpansion
{\displaystyle {d\bar{R}_{R}(u) \over du}}%
%EndExpansion
=-(1+\bar{R}_{R}(u))%
%TCIMACRO{\dfrac{2\gamma \sqrt{1-\beta ^{2}}}{i\omega } }
%BeginExpansion
{\displaystyle {2\gamma \sqrt{1-\beta ^{2}} \over i\omega }}%
%EndExpansion
+\bar{R}_{R}(u)\left( \beta f^{\prime }-\beta +1+f^{\prime }\right)
\label{76b}
\end{equation}

\noindent The fact that we end up with a differential equation is a direct
result of the derivatives acting on $R,T.$ To solve it we should express all
quantities appearing in the right hand side of (\ref{76b}) in terms of $u$.
For a specific trajectory $g(t)$ the functions $\beta (u),$ $f^{\prime }(u)$
are calculable along the lines of Appendix A (at least in principle).
Equation (\ref{76b}) is a linear first order differential equation and a
general solution may be readily written down, however it is hardly
illuminating. The study of a specific trajectory will not be pursued in the
present note.

Equation (\ref{76b}) for $\bar{R}_{R}$ may be expressed using $\beta $
rather than $u$ as the independent variable. To this end we write, using (%
\ref{208}), 
\[
\beta f^{\prime }-\beta +1+f^{\prime }=2(1-\beta ) 
\]

\noindent and rewrite (\ref{76b}) in the form 
\begin{equation}
%TCIMACRO{\dfrac{1}{i\omega } }
%BeginExpansion
{\displaystyle {1 \over i\omega }}%
%EndExpansion
%TCIMACRO{\dfrac{d\bar{R}_{R}(u)}{du} }
%BeginExpansion
{\displaystyle {d\bar{R}_{R}(u) \over du}}%
%EndExpansion
=-(1+\bar{R}_{R}(u))%
%TCIMACRO{\dfrac{\gamma \sqrt{1-\beta ^{2}}}{i\omega } }
%BeginExpansion
{\displaystyle {\gamma \sqrt{1-\beta ^{2}} \over i\omega }}%
%EndExpansion
+\bar{R}_{R}(u)\left( 1-\beta \right)  \label{203}
\end{equation}

\noindent We can use (\ref{211}) to write the above equation in the form 
\begin{equation}
%TCIMACRO{\dfrac{a}{i\omega (1-\beta )} }
%BeginExpansion
{\displaystyle {a \over i\omega (1-\beta )}}%
%EndExpansion
%TCIMACRO{\dfrac{d\bar{R}_{R}(\beta )}{d\beta } }
%BeginExpansion
{\displaystyle {d\bar{R}_{R}(\beta ) \over d\beta }}%
%EndExpansion
=-(1+\bar{R}_{R}(\beta ))%
%TCIMACRO{\dfrac{\gamma \sqrt{1-\beta ^{2}}}{i\omega } }
%BeginExpansion
{\displaystyle {\gamma \sqrt{1-\beta ^{2}} \over i\omega }}%
%EndExpansion
+\bar{R}_{R}(\beta )\left( 1-\beta \right)  \label{213}
\end{equation}

\noindent For a specific trajectory the acceleration $a$ may be expressed as
a function of the velocity $\beta $.

As a test of the algebra and as an illustration let us consider the case of
uniform motion. Then $\beta $ is constant, $f^{\prime }$ is read off (\ref
{e9b}), and (\ref{76b}) admits the $u$-independent solution 
\begin{equation}
\bar{R}_{R}^{B}(\omega )\equiv 
%TCIMACRO{
%\dfrac{-i\gamma }{i\gamma +\dfrac{\omega (1-B)}{\gamma \sqrt{1-B^{2}}}} }
%BeginExpansion
{\displaystyle {-i\gamma  \over i\gamma +{\displaystyle {\omega (1-B) \over \gamma \sqrt{1-B^{2}}}}}}%
%EndExpansion
\label{59}
\end{equation}

\noindent This is the $R$ of (\ref{rt}) as expected, the quantity $\omega
(1-B)/\sqrt{1-B^{2}}$ being the photon frequency in the frame of the mirror
(cf (\ref{59c})).

Consider a mirror moving at uniform velocity $\beta _{0}$ till it reaches
the origin {\it O}, accelerates from {\it O} to {\it P}, and then continues
at uniform velocity $\beta _{P}$. The differential equation must then be
solved from $\beta =\beta _{0}$ to $\beta =\beta _{P}$ with initial
condition 
\begin{equation}
\bar{R}_{R}(\omega ,\beta )=\bar{R}_{R}^{\beta _{0}}(\omega )  \label{214}
\end{equation}

\noindent where the right hand side of (\ref{214}) is given by (\ref{59}).
At point {\it P }the solution of the differential equation joins smoothly
with $\bar{R}_{R}^{\beta _{P}}(\omega )$.

(ii)\ Modes transmitted to the right (equations (\ref{72}), (\ref{73})): 
\begin{equation}
%TCIMACRO{\dfrac{2}{i\omega } }
%BeginExpansion
{\displaystyle {2 \over i\omega }}%
%EndExpansion
%TCIMACRO{\dfrac{d\bar{R}_{L}(v)}{dv} }
%BeginExpansion
{\displaystyle {d\bar{R}_{L}(v) \over dv}}%
%EndExpansion
=(1+\bar{R}_{L}(v))%
%TCIMACRO{\dfrac{2\gamma \sqrt{1-\beta ^{2}}}{i\omega } }
%BeginExpansion
{\displaystyle {2\gamma \sqrt{1-\beta ^{2}} \over i\omega }}%
%EndExpansion
+\bar{R}_{L}(v)\left( \beta p^{\prime }-p^{\prime }-\beta -1\right)
\label{77b}
\end{equation}

\noindent Now everything on the right hand side should be expressed in terms
of $v$. This again is in principle straightforward once the equation of the
trajectory is given.

We can repeat the analysis of (i) above. We use (\ref{207}) to write 
\[
\beta p^{\prime }-p^{\prime }-\beta -1=-2(1+\beta ) 
\]

\noindent and rewrite (\ref{77b}) in the form 
\begin{equation}
%TCIMACRO{\dfrac{1}{i\omega } }
%BeginExpansion
{\displaystyle {1 \over i\omega }}%
%EndExpansion
%TCIMACRO{\dfrac{d\bar{R}_{L}(v)}{dv} }
%BeginExpansion
{\displaystyle {d\bar{R}_{L}(v) \over dv}}%
%EndExpansion
=(1+\bar{R}_{L}(v))%
%TCIMACRO{\dfrac{\gamma \sqrt{1-\beta ^{2}}}{i\omega } }
%BeginExpansion
{\displaystyle {\gamma \sqrt{1-\beta ^{2}} \over i\omega }}%
%EndExpansion
-\bar{R}_{L}(v)\left( 1+\beta \right)  \label{215}
\end{equation}

\noindent We use (\ref{212}) to rewrite the above in the form 
\begin{equation}
%TCIMACRO{\dfrac{a}{i\omega (1+\beta )} }
%BeginExpansion
{\displaystyle {a \over i\omega (1+\beta )}}%
%EndExpansion
%TCIMACRO{\dfrac{d\bar{R}_{L}(\beta )}{d\beta } }
%BeginExpansion
{\displaystyle {d\bar{R}_{L}(\beta ) \over d\beta }}%
%EndExpansion
=(1+\bar{R}_{L}(\beta ))%
%TCIMACRO{\dfrac{\gamma \sqrt{1-\beta ^{2}}}{i\omega } }
%BeginExpansion
{\displaystyle {\gamma \sqrt{1-\beta ^{2}} \over i\omega }}%
%EndExpansion
-\bar{R}_{L}(\beta )\left( 1+\beta \right)  \label{216}
\end{equation}

Concerning the solution of (\ref{216}), the reader is referred to the
remarks made following (\ref{213}). In the case of uniform motion we obtain
in a similar way as before the solution 
\begin{equation}
\bar{R}_{L}^{B}(\omega )\equiv 
%TCIMACRO{
%\dfrac{-i\gamma }{i\gamma +\dfrac{\omega (1+B)}{\gamma \sqrt{1-B^{2}}}} }
%BeginExpansion
{\displaystyle {-i\gamma  \over i\gamma +{\displaystyle {\omega (1+B) \over \gamma \sqrt{1-B^{2}}}}}}%
%EndExpansion
\label{75}
\end{equation}

\noindent Notice that in the denominator we again have the frequency in the
mirror's rest frame. Notice also the expected difference in the sign of $%
\beta $ between equations (\ref{59}) and (\ref{75}). This is due to the fact
that the transmitted wave is now in the opposite direction than before.
Similarly we obtain differential equations for the last two classes of modes.

(iii) Modes incident from the left (equations (\ref{54}), (\ref{55})): 
\begin{equation}
%TCIMACRO{\dfrac{2}{i\omega } }
%BeginExpansion
{\displaystyle {2 \over i\omega }}%
%EndExpansion
%TCIMACRO{\dfrac{dT_{L}(u)}{du} }
%BeginExpansion
{\displaystyle {dT_{L}(u) \over du}}%
%EndExpansion
=-T_{L}(u)%
%TCIMACRO{\dfrac{2\gamma \sqrt{1-\beta ^{2}}}{i\omega } }
%BeginExpansion
{\displaystyle {2\gamma \sqrt{1-\beta ^{2}} \over i\omega }}%
%EndExpansion
+T_{L}(u)\left( \beta f^{\prime }-\beta +1+f^{\prime }\right) -\left( \beta
f^{\prime }-\beta +1+f^{\prime }\right)  \label{58b}
\end{equation}

(iv) Modes incident from the right (equations (\ref{52}), (\ref{53})):

\noindent

\begin{equation}
%TCIMACRO{\dfrac{2}{i\omega } }
%BeginExpansion
{\displaystyle {2 \over i\omega }}%
%EndExpansion
%TCIMACRO{\dfrac{dT_{R}(v)}{dv} }
%BeginExpansion
{\displaystyle {dT_{R}(v) \over dv}}%
%EndExpansion
=-T_{R}(v)%
%TCIMACRO{\dfrac{2\gamma \sqrt{1-\beta ^{2}}}{i\omega } }
%BeginExpansion
{\displaystyle {2\gamma \sqrt{1-\beta ^{2}} \over i\omega }}%
%EndExpansion
+T_{R}(v)\left( \beta p^{\prime }-p^{\prime }-\beta -1\right) -\left( \beta
p^{\prime }-p^{\prime }-\beta -1\right)  \label{74b}
\end{equation}

Equations (\ref{58b}), (\ref{74b}) can be cast in terms of $a$ and $\beta $
via the same manipulations used in obtaining (\ref{213}) and (\ref{216}).
The four uncoupled linear first order differential equations (\ref{76b}), (%
\ref{77b}), (\ref{58b}), and (\ref{74b}) may be solved (at least to some
approximation) once the form of the trajectory is specified.

After we obtain the solutions, or an approximation to them, we can use the
formalism of the preceding sections to calculate the Bogolubov amplitudes.
There are some obvious changes. The indices $i$, $j$ appearing in (\ref{e011}%
), (\ref{e0011}) now stand for both the frequency $\omega $ and the
direction of incidence ($R$ or $L$). The emission amplitude $\beta \left(
\omega _{L,R},\omega _{L,R}^{\prime }\right) $ is given by the standard
expression (\ref{e11}) where now the integration ranges from $z=-\infty $ to 
$\infty $.

The covariant model of a dispersive mirror allows estimates about the
spectrum of the emitted radiation without solving the above differential
equations. The amplitude $\beta \left( \omega ,\omega ^{\prime }\right) $
calculated in the previous section in the context of perfect mirrors
corresponds to $\beta \left( \omega _{R},\omega _{R}^{\prime }\right) $ in
the notation of this section. If we describe photons in the rest frame of
the mirror then roughly speaking frequencies less than $\gamma $ are
reflected whereas the ones greater than $\gamma $ are transmitted (this is
no more than a very crude qualitative estimate resulting from (\ref{rt})).
According to the treatment in the preceding sections $\omega ^{\prime }$
labels the {\it in} states where the mirror is at rest. Thus the {\it in }%
frequencies that roughly matter are in the range $\omega ^{\prime }<\gamma $
. The frequency $\omega $ labels an {\it out} state and is connected to the
frequency $\Omega ^{\prime }$ in the mirror's rest frame via (\ref{e10}) or (%
\ref{59c}) depending on the direction of incidence, both essentially
amounting to a Lorentz transformation from one frame to another. For a
receding mirror and right incident photons the two frequencies are connected
in terms of $\beta _{P}$ (the velocity at {\it P}$)$ 
\begin{equation}
\Omega ^{\prime }=\omega \sqrt{%
%TCIMACRO{\dfrac{1+\left| \beta _{P}\right| }{1-\left| \beta _{P}\right| } }
%BeginExpansion
{\displaystyle {1+\left| \beta _{P}\right|  \over 1-\left| \beta _{P}\right| }}%
%EndExpansion
}  \label{c10}
\end{equation}

\noindent Hence the {\it out }frequencies that matter for $\beta \left(
\omega _{R},\omega _{R}^{\prime }\right) $ are in the range 
\begin{equation}
\omega <\gamma \sqrt{%
%TCIMACRO{\dfrac{1-\left| \beta _{P}\right| }{1+\left| \beta _{P}\right| } }
%BeginExpansion
{\displaystyle {1-\left| \beta _{P}\right|  \over 1+\left| \beta _{P}\right| }}%
%EndExpansion
}  \label{c11}
\end{equation}
This limit is easily saturated in the case of high terminal velocity $\beta
_{P}$. This of course does not apply to the amplitude $\beta (\omega
_{L}^{\prime },\omega _{L})$ for a mirror executing the same motion where
the signs in (\ref{c10}), (\ref{c11}) are interchanged between numerator and
denominator. Having made these qualitative remarks one must be warned that
the role of high frequencies (compared to $\gamma $) in the context of
dispersive mirrors should not be underestimated; experience has shown that a
finite $\gamma $ does not prevent integrations from going ultraviolet
divergent.

\section{Conclusions}

We considered a perfect mirror starting from rest and accelerating along the
trajectory $g(t)=-\ln (\cosh t)$ till it reaches a speed arbitrarily close
to $c$ and then continuing at uniform velocity. We presented an exact
evaluation of the Bogolubov amplitude $\beta \left( \omega ,\omega ^{\prime
}\right) $ and showed that $\left| \beta \left( \omega ,\omega ^{\prime
}\right) \right| ^{2}$ goes as $\left( \omega ^{\prime }\right) ^{-5}$, this
being a result of the velocity staying continuous throughout the trajectory
and of the fact that the trajectory is asymptotically inertial. We stress
that the problem examined here is not the problem studied by Fulling and \
Davies (1976) and Davies and Fulling (1977) where the mirror accelerates
along $g(t)=-\ln (\cosh t)$ ad infinitum. Observe that the $\beta \left(
\omega ,\omega ^{\prime }\right) $ amplitude for the Fulling and Davies
problem cannot be obtained as the limit $r\rightarrow \ln 2$ of the
amplitude evaluated in this note. This is immediately demonstrated by the
different behaviour for large $\omega ^{\prime }$ (in the DF problem $\left|
\beta \left( \omega ,\omega ^{\prime }\right) \right| ^{2}$ goes
asymptotically as $1/\omega ^{\prime }$). There is thus no question of the
thermal spectrum arising in the case of an asymptotically inertial
trajectory, regardless of how close to the speed of light the final velocity
lies. The mirror trajectory $g(t)=-\ln \left( \cosh t\right) $ is of
interest as a model of black hole collapse. However a realistic mirror (or
any other interface) is not expected to have its velocity forever
increasing. The results of this paper can thus serve as a guide to what one
should expect in the realistic case. In section 4 we constructed the normal
modes for a double faced dispersive mirror accelerating to relativistic
velocities. The formalism of the previous sections can then be applied in
the calculation of the Bogolubov amplitudes in the dispersive case. This may
be of relevance to the calculation of quasiparticle emission from an
accelerating He3 A-B interface.

Note: After the completion of the work in section 4 the paper by Nicolaevici
(2001) appeared where the model of Barton and Calogeracos (1995) is used to
obtain the differential equations for the $T$ and $R$ amplitudes. The author
focuses on the calculation of local field quantities rather than on the
calculation of Bogolubov amplitudes.

\noindent \noindent {\bf Acknowledgments.}

The author is indebted to S\ A\ Fulling, G\ Plunien, R\ Schutzhold, and G\ E
Volovik for discussions and/or \noindent correspondence. He also wishes to
thank M\ Paalanen and the Low Temperature Laboratory, Helsinki University of
Technology, and G\ Soff and the Institute for Theoretical Physics, Dresden
Technical University for their hospitality.\bigskip
\newpage\
\noindent {\bf Appendix A: The trajectory} $z$=-$\frac{1}{\kappa }\ln \left(
\cosh \kappa t\right) $.

A spacetime point is labelled by two coordinates $\left( u,v\right) $ and
the trajectory is specified by the function $u=f(v)$ or equally well by its
inverse $v=p(u).$ From the first of (\ref{e01}) and the equation of the
trajectory we get after some trivial algebra 
\begin{equation}
t=%
%TCIMACRO{\dfrac{1}{2\kappa } }
%BeginExpansion
{\displaystyle {1 \over 2\kappa }}%
%EndExpansion
\ln \left( 2e^{\kappa u}-1\right)  \label{a1}
\end{equation}

\noindent The velocity is 
\[
\beta (t)=%
%TCIMACRO{\dfrac{dz}{dt}}
%BeginExpansion
{\displaystyle {dz \over dt}}%
%EndExpansion
=-\tanh \kappa t 
\]

\noindent and 
\[
\sqrt{1-\beta ^{2}}=\frac{1}{\cosh \kappa t} 
\]
The acceleration is 
\begin{equation}
a=\frac{d\beta }{dt}=-\frac{\kappa }{\cosh ^{2}\kappa t}  \label{a6}
\end{equation}

\noindent and the proper acceleration is 
\begin{equation}
\alpha =a\left( 1-\beta ^{2}\right) ^{-3/2}=-\kappa \cosh \kappa t
\label{a7}
\end{equation}

Equations (\ref{e04}) and (\ref{a1}) determine $p(u):$

\noindent 
\begin{equation}
p(u)=\frac{1}{\kappa }\ln \left( 2e^{\kappa u}-1\right) -u  \label{a3}
\end{equation}

\noindent

\noindent We also quote the derivative 
\begin{equation}
p^{\prime }(u)=\frac{1}{2e^{\kappa u}-1}  \label{a8}
\end{equation}
\noindent The inverse to $p(u)$ is 
\begin{equation}
f(v)=-\frac{1}{\kappa }\ln \left( 2-e^{\kappa v}\right)  \label{a4}
\end{equation}

\noindent Relations (\ref{a3}), (\ref{a4}) are valid throughout the
accelerated trajectory. Obviously the trajectory asymptotically approaches
the null line $l1$ (for $\kappa =1$) (figure 7)$.$ Null lines lying above it
(like $l2$) never meet the trajectory. This is reflected by the fact that $f$
is not real for $v\geq \left( \ln 2\right) /\kappa $.

To determine the velocity we use (\ref{207}) to obtain 
\begin{equation}
\beta (u)=e^{-\kappa u}-1  \label{a5}
\end{equation}

\noindent (approaching $-1$ as expected when $v\rightarrow \frac{\ln 2}{%
\kappa }-$). From (\ref{208}) we determine the velocity as a function of $v$%
\begin{equation}
\beta (v)=1-e^{\kappa v}  \label{a2}
\end{equation}

\newpage\
%%%%%%%%%%%%%%%%%%%%%%%%%%%%%%%%%%%%%%%%%%%%%%%%%%%%%%%%%%%%%%%%%%%%%%%%%%%%%
\begin{figure}[tbph]
%h=here,t=top,b=bottom,p=separate figure page
\par
\begin{center}
\leavevmode
\includegraphics[width=0.7\linewidth]{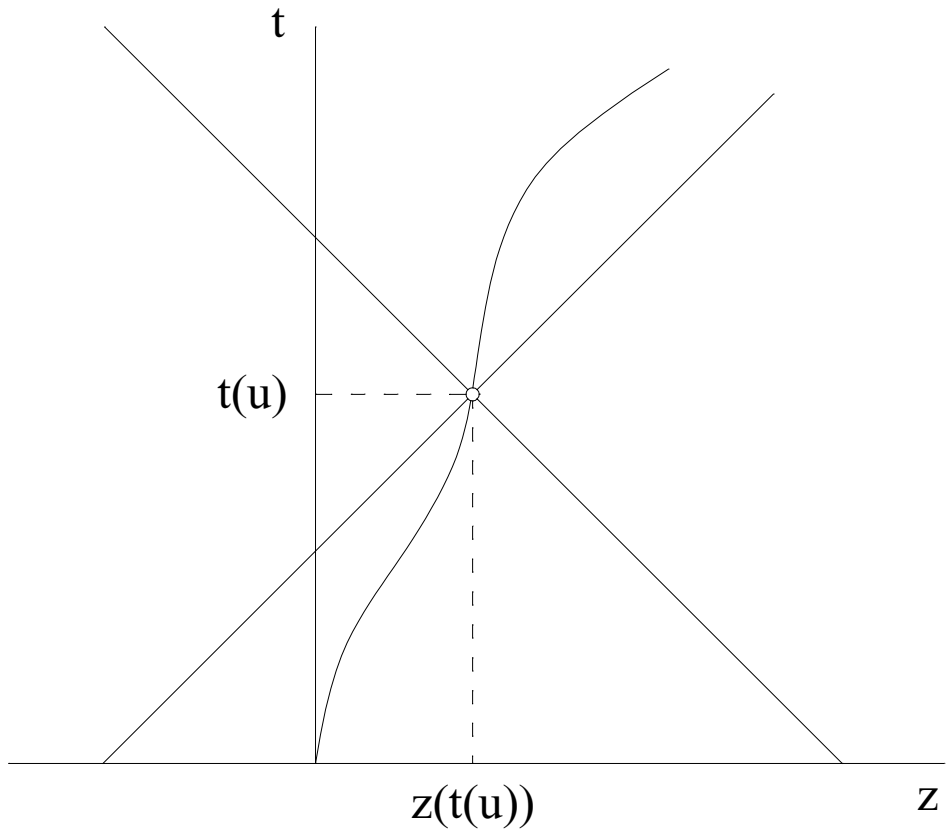} \medskip
\end{center}
\caption{The spacetime coordinates $t,z$ of a point on the trajectory as functions of 
$u$}                                                                   
\end{figure}
%%%%%%%%%%%%%%%%%%%%%%%%%%%%%%%%%%%%%%%%%%%%%%%%%%%%%%%%%%%%%%%%%%%%%%%%%%
\newpage\
%%%%%%%%%%%%%%%%%%%%%%%%%%%%%%%%%%%%%%%%%%%%%%%%%%%%%%%%%%%%%%%%%%%%%%%%%%%%%
\begin{figure}[tbph]
%h=here,t=top,b=bottom,p=separate figure page
\par
\begin{center}
\leavevmode
\includegraphics[width=0.7\linewidth]{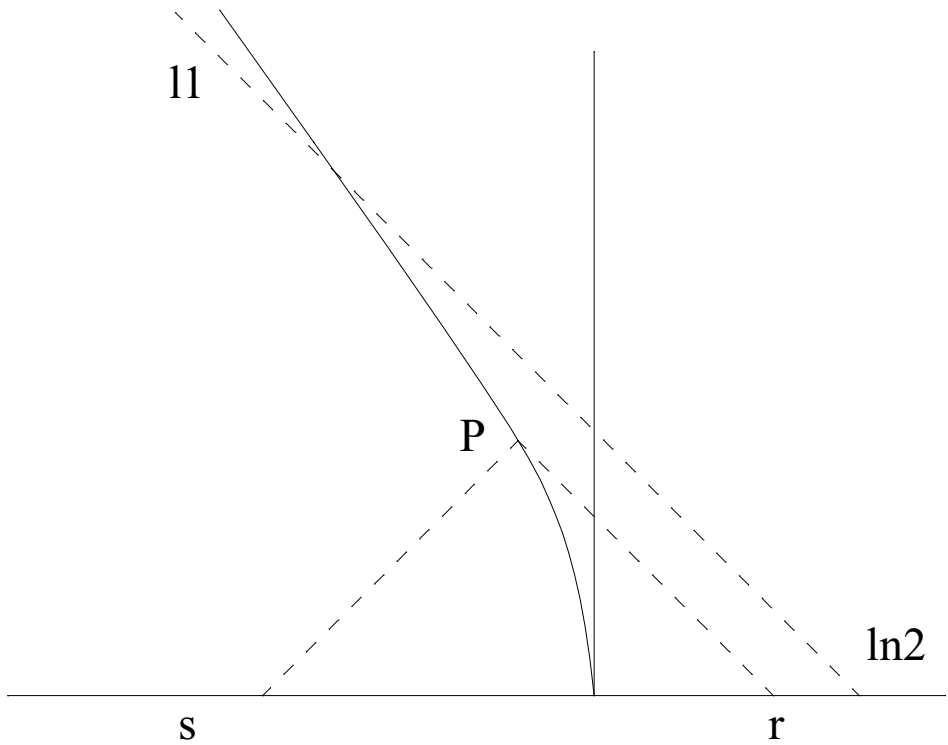} \medskip
\end{center}
\caption{Asymptotically inertial trajectory starting from rest at 
{\it O, }accelerating from {\it O} to {\it P}, and reverting to uniform
acceleration at {\it P.}}
\end{figure}
%%%%%%%%%%%%%%%%%%%%%%%%%%%%%%%%%%%%%%%%%%%%%%%%%%%%%%%%%%%%%%%%%%%%%%%%%%
\newpage\
%%%%%%%%%%%%%%%%%%%%%%%%%%%%%%%%%%%%%%%%%%%%%%%%%%%%%%%%%%%%%%%%%%%%%%%%%%%%%
\begin{figure}[tbph]
%h=here,t=top,b=bottom,p=separate figure page
\par
\begin{center}
\leavevmode
\includegraphics[width=0.7\linewidth]{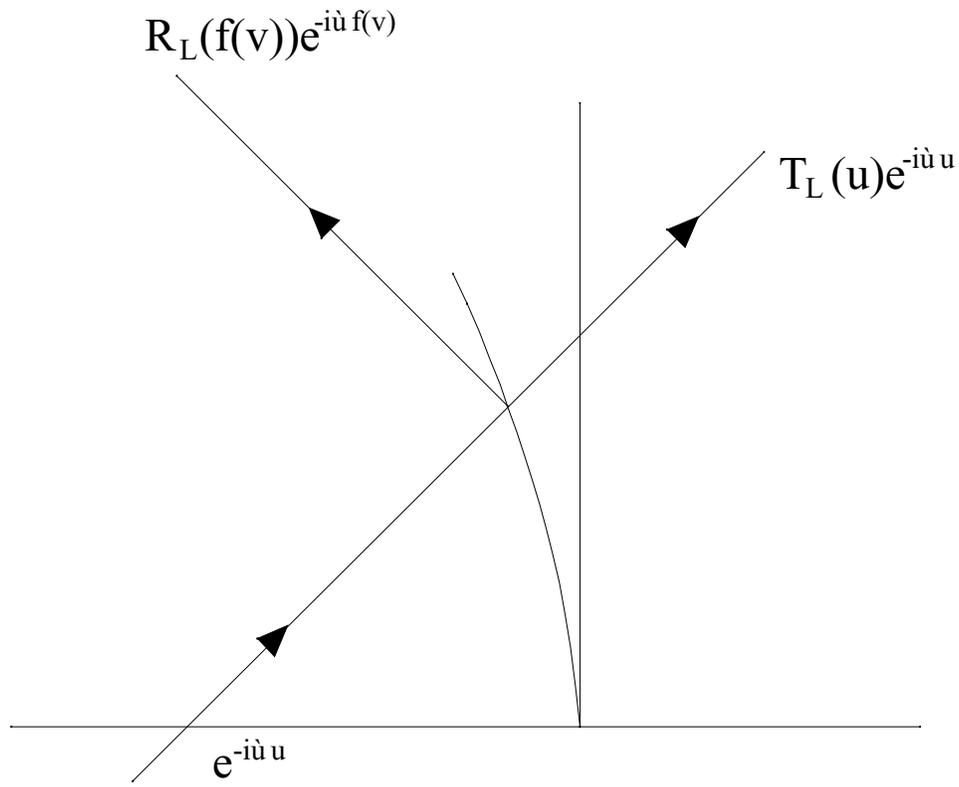} \medskip
\end{center}
\caption{{\it In} mode with  $\exp \left( -i\omega u\right) $ wave
incident from left, transmitted wave $T_{L}(u)\exp \left( -i\omega u\right) $
to the right, reflected wave $R_{L}(f(v))\exp (-i\omega f(v))$ to the left}
\end{figure}
%%%%%%%%%%%%%%%%%%%%%%%%%%%%%%%%%%%%%%%%%%%%%%%%%%%%%%%%%%%%%%%%%%%%%%%%%%
\newpage\
%%%%%%%%%%%%%%%%%%%%%%%%%%%%%%%%%%%%%%%%%%%%%%%%%%%%%%%%%%%%%%%%%%%%%%%%%%%%%
\begin{figure}[tbph]
%h=here,t=top,b=bottom,p=separate figure page
\par
\begin{center}
\leavevmode
\includegraphics[width=0.7\linewidth]{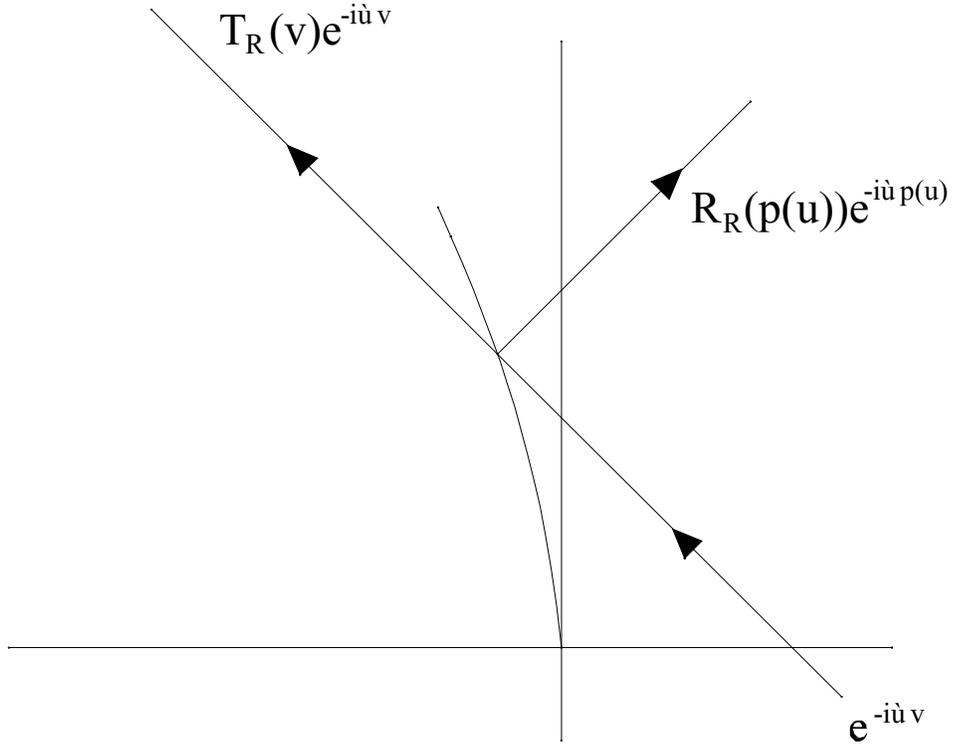} \medskip
\end{center}
\caption{{\it In} mode with $\exp \left( -i\omega v\right) $ wave
incident from right, transmitted wave  $T_{R}(v)\exp \left( -i\omega
v\right) $ to the left, reflected wave $R_{R}(p(u))\exp (-i\omega p(u))$ to
the right}
\end{figure}
%%%%%%%%%%%%%%%%%%%%%%%%%%%%%%%%%%%%%%%%%%%%%%%%%%%%%%%%%%%%%%%%%%%%%%%%%%
\newpage\
%%%%%%%%%%%%%%%%%%%%%%%%%%%%%%%%%%%%%%%%%%%%%%%%%%%%%%%%%%%%%%%%%%%%%%%%%%%%%
\begin{figure}[tbph]
%h=here,t=top,b=bottom,p=separate figure page
\par
\begin{center}
\leavevmode
\includegraphics[width=0.7\linewidth]{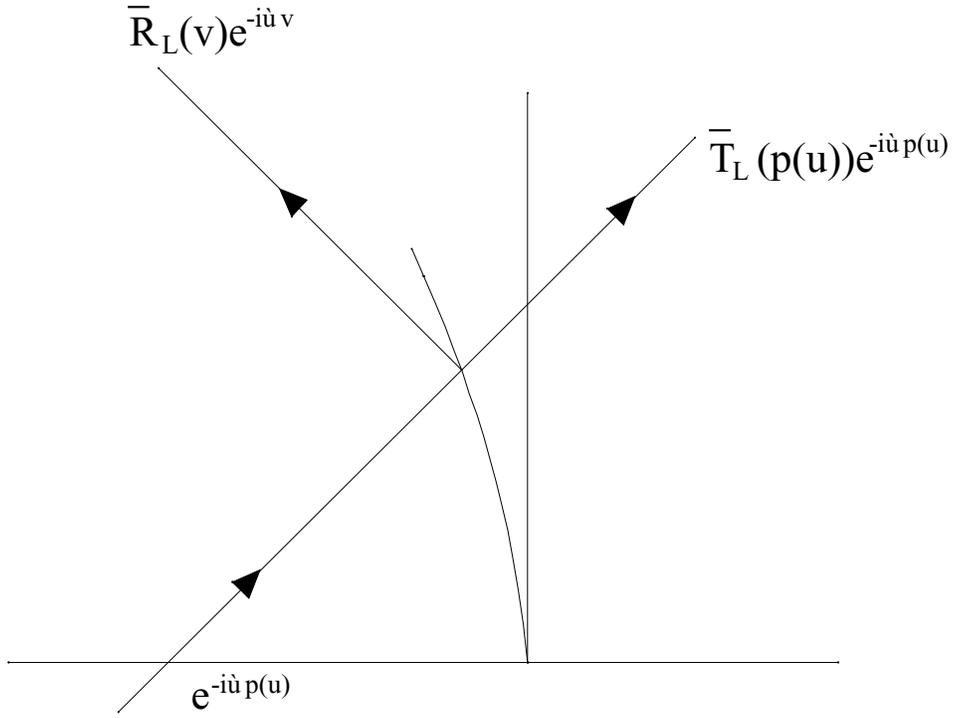} \medskip
\end{center}
\caption{{\it Out }mode with  $\overline{R}_{L}(v)\exp (-i\omega v)$
wave to the left , $\overline{T}_{L}(p(u))\exp (-i\omega p(u))$ wave to the
right, back scattered wave $\exp (-i\omega p(u))$}
\end{figure}
%%%%%%%%%%%%%%%%%%%%%%%%%%%%%%%%%%%%%%%%%%%%%%%%%%%%%%%%%%%%%%%%%%%%%%%%%%
\newpage\
%%%%%%%%%%%%%%%%%%%%%%%%%%%%%%%%%%%%%%%%%%%%%%%%%%%%%%%%%%%%%%%%%%%%%%%%%%%%%
\begin{figure}[tbph]
%h=here,t=top,b=bottom,p=separate figure page
\par
\begin{center}
\leavevmode
\includegraphics[width=0.7\linewidth]{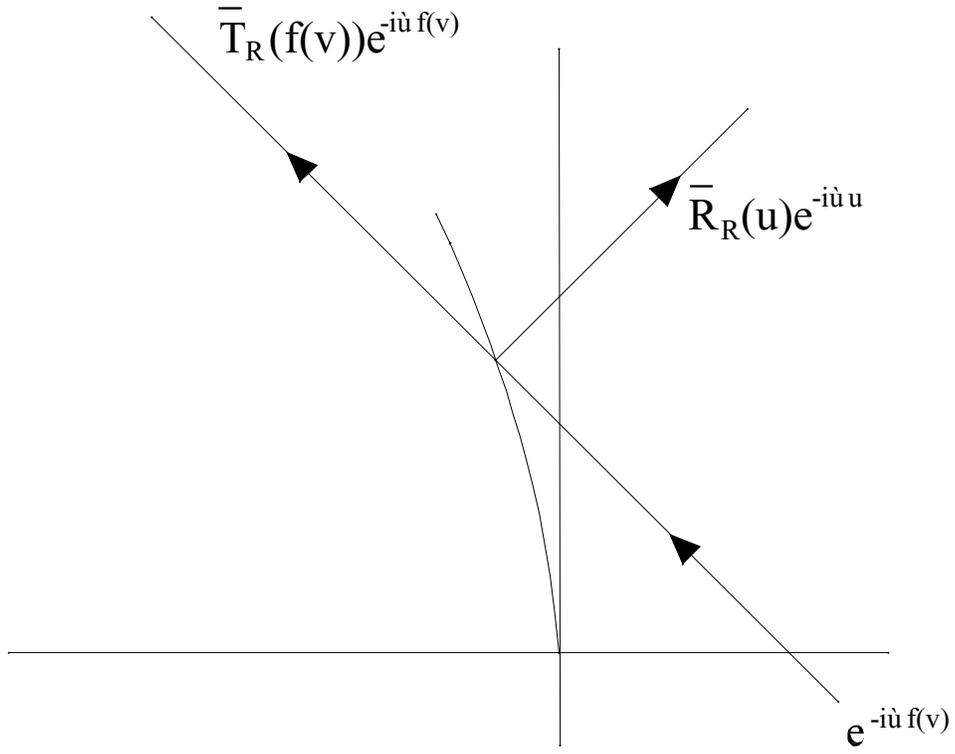} \medskip
\end{center}
\caption{{\it Out }mode with $\overline{T}_{R}(f(v))\exp (-i\omega
f(v))$ wave to the left , $\overline{R}_{R}(u)\exp (-i\omega u)$ wave to the
right, back scattered wave $\exp (-i\omega f(v))$}
\end{figure}
%%%%%%%%%%%%%%%%%%%%%%%%%%%%%%%%%%%%%%%%%%%%%%%%%%%%%%%%%%%%%%%%%%%%%%%%%%
\newpage\
%%%%%%%%%%%%%%%%%%%%%%%%%%%%%%%%%%%%%%%%%%%%%%%%%%%%%%%%%%%%%%%%%%%%%%%%%%%%%
\begin{figure}[tbph]
%h=here,t=top,b=bottom,p=separate figure page
\par
\begin{center}
\leavevmode
\includegraphics[width=0.7\linewidth]{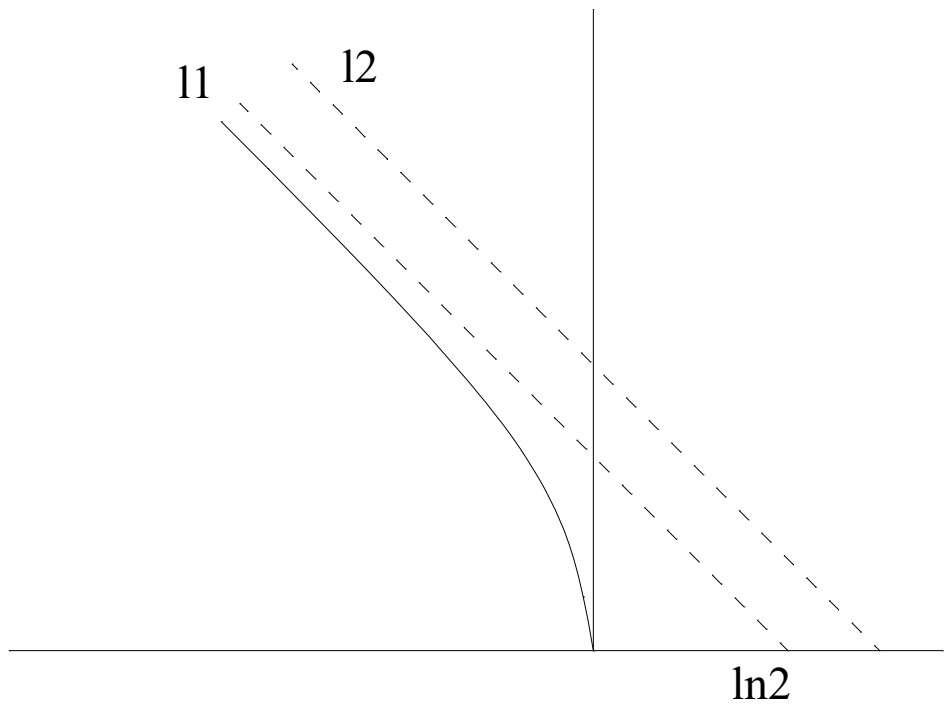} \medskip
\end{center}
\caption{The asymptote to the trajectory $g(t)=-\ln \left( \cosh
t\right) $ and a null line lying above it}
\end{figure}
%%%%%%%%%%%%%%%%%%%%%%%%%%%%%%%%%%%%%%%%%%%%%%%%%%%%%%%%%%%%%%%%%%%%%%%%%%

\end{document}